\documentclass[aps]{revtex4-1}

\usepackage{graphicx}
\usepackage{amsmath}
\usepackage{amssymb}
\usepackage{color}
\usepackage{setspace}

\begin{document}

\author{J. Scheibert$^1$}
\affiliation{$^1$Laboratory of Tribology and System Dynamics, UMR CNRS-ECL 5513, University of Lyon, France}
\author{O. Galland$^2$}
\affiliation{$^2$Physics of Geological Processes (PGP), Department of Geosciences, University of Oslo, Norway}
\author{A. Hafver$^2$}
\affiliation{$^2$Physics of Geological Processes (PGP), Department of Geosciences, University of Oslo, Norway}

\title{\textcolor{black}{Inelastic deformation during sill and laccolith emplacement: insights from an analytic elasto-plastic model}}

\begin{abstract}
\textcolor{black}{Numerous geological observations evidence that inelastic deformation occurs during sills and laccoliths emplacement. However, most models of sill and laccolith emplacement neglect inelastic processes by assuming purely elastic deformation of the host rock. This assumption has never been tested, so that the role of inelastic deformation on the growth dynamics of magma intrusions remains poorly understood.} In this paper, we introduce the first analytical model of shallow sill and laccolith emplacement that accounts for elasto-plastic deformation of the host rock. It considers the intrusion's overburden as a thin elastic bending plate attached to an elastic-perfectly-plastic foundation. We find that, for geologically realistic values of the model parameters, the horizontal extent of the plastic zone $l_p$ is much smaller than the \textcolor{black}{radius} of the intrusion $a$. By modeling the quasi-static growth of a sill, we find that the ratio $l_p/a$ decreases during propagation, as $1 / \sqrt{a^4 \Delta P}$, with $\Delta P$ the magma overpressure. The model also shows that the extent of the plastic zone decreases with the intrusion's depth, while it increases if the host rock is weaker. \textcolor{black}{Comparison between our elasto-plastic model and existing purely elastic models shows that plasticity can have a significant effect on intrusion propagation dynamics, with \textit{e.g.} up to a doubling of the overpressure necessary for the sill to grow.} Our results suggest that plasticity effects might be small for large sills, but conversely that they might be substantial for early sill propagation.
\end{abstract}

\maketitle





\section{Introduction}

Over the past few decades, geological field studies \cite{Polteau2008, Galerne2011, Schofield2012} and seismic reflection data \cite{Hansen2004, Planke2005, Hansen2008, Polteau2008, Galland2009, Galerne2011, Magee2014, Magee2016} have revealed the presence of voluminous igneous complexes in sedimentary basins worldwide. Igneous intrusions in these basins exhibit various shapes, from flat or saucer-shaped sills, to laccoliths \cite{Planke2005, Jackson2013}. It has been demonstrated that intrusive rocks and processes have major impacts on the thermal and structural evolutions of sedimentary basins \cite{Petford2003, Schutter2003}. Among others (1) sills provide heat that locally maturates the organic matter in the surrounding sediments \cite{Svensen2004, Rodriguezmonreal2009, Aarnes2011}, (2) sill emplacement may cause uplift and deformation of the host rock, forming broad domes, or forced folds, of their overlaying strata \cite{Jackson1990, Trude2003, Hansen2006, Jackson2013, Agirrezabala2015}, and (3) damage induced by the emplacement of magma produces fractures in the host rock that enhance fluid flow \cite{Delaney1981, Meriaux1999, Chevallier2004, Senger2015}. 

\textcolor{black}{Sills also represent significant parts of the plumbing systems of active volcanoes worldwide. Field studies have highlighted the presence of sills and laccoliths in volcanic complexes \cite[e.g., ][]{Pasquare2007, Burchardt2008}. Numerous geodetic surveys have also revealed the emplacement of sills, some of which resulting in eruptions, among others, in the Galápagos Islands \cite[e.g.,][]{Amelung2000}, Eyjafjallaj{\"o}kull volcano, Iceland \cite[e.g.,][]{Pedersen2004, Pedersen2006, Sigmundsson2010}, in the Afar region, Ethiopia \cite[e.g.,][]{Nobile2012, Pagli2012}, and Piton de la Fournaise volcano, R\'{e}union Island \cite[e.g.,][]{Chaput2014}.}

In sedimentary basins, existing theoretical and numerical models of sill and laccolith emplacement account for elastic host rock only. Classical models, \textcolor{black}{as well as very recent ones,} consider the sill overburden as an elastic thin plate clamped to a perfectly rigid basement \cite{Pollard1973, Jackson1990, Scaillet1995, Kerr1998, Goulty2008, Bunger2011, Michaut2011, Michaut2014, Thorey2014, Michaut2016}, and assume intrusion propagation to obey Linear Elastic Fracture Mechanics (LEFM) theory. Because these models are clamped, they only account for deformation above the intrusion, which is not realistic \cite[][and references therein]{Galland2013}. To overcome this limitation, a more advanced mathematical formulation considers a thin elastic plate on top of a deformable elastic foundation \cite{Kerr1998, Galland2013}. \textcolor{black}{The latter models produce realistic elastic deformation of sills overburden \cite[see discussion by][]{Agirrezabala2015} ; however, they are also limited to purely elastic propagation of the intrusions.}

\begin{figure}
\begin{center}
\includegraphics[width=0.8\columnwidth]{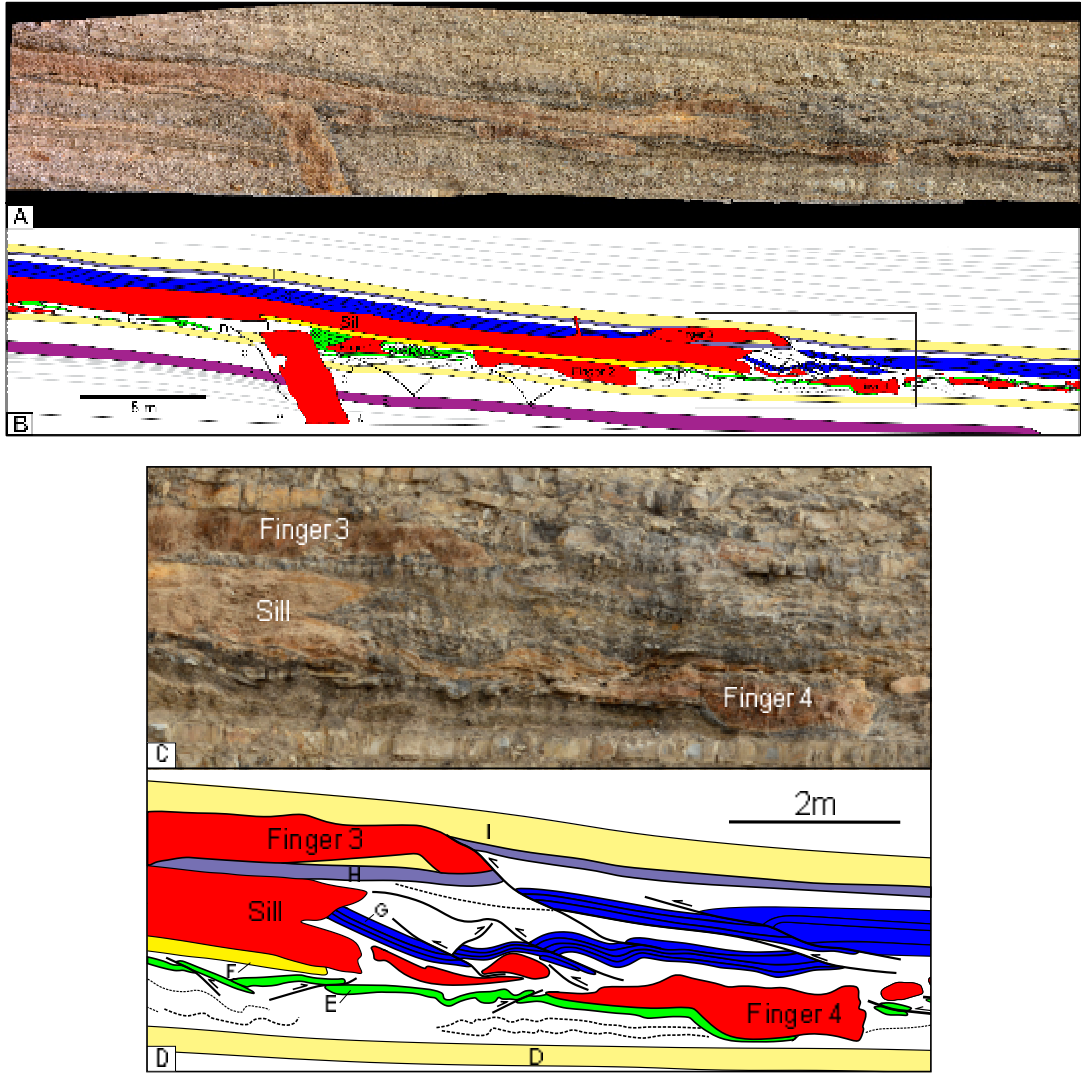}
\caption{Field ortho-rectified image (A) and interpreted drawing (B) of outcrop exposing a sheet-like sill, magmatic fingers and the associated structures in the shale-carbonate host rock, Cuesta del Chihuido, Mendoza Province, Argentina \cite{SpacapanSubmitted}. The box locates the zoomed image (C) and associated interpreted drawing (D). The outcrop shows that the sill tip is not sharp, and that substantial inelastic deformation (brittle faulting of the carbonate layers, ductile flow of the shale layers) accommodates the emplacement and propagation of the sill. Detailed descriptions of the structures and associated mechanisms can be found is \cite{SpacapanSubmitted}.}
\label{figure1}
\end{center}
\end{figure}

\textcolor{black}{\cite{Rubin1993} argues that the fracture toughness propagation criterion used in LEFM theory does not apply for intrusions deeper than a few hundred meters (\textit{i.e.} for most sills and laccoliths). In addition, recent geological and geophysical observations show that some inelastic deformation accommodates sill and laccolith emplacement in sedimentary formations (Fig.~\ref{figure1}). At shallow levels, igneous sills often intrude into rocks that deform inelastically, such as soft shale formations \cite[e.g.,][]{Planke2005, Jackson2013, SpacapanSubmitted}. \cite{Pollard1975}, \cite{Duffield1986}, \cite{Schofield2012, Schofield2014} and \cite{SpacapanSubmitted} provide field evidence that inelastic deformation in the vicinity of intrusion tips might play a significant role in the emplacement of sills and dikes in soft rock formations. Such inelastic deformation involves, among others, joints and micro-fractures \cite{Delaney1981} and brittle and ductile faulting \cite{Pollard1973b, Pollard1975, SpacapanSubmitted}.}

\textcolor{black}{In active volcanoes, geodetic measurements are commonly interpreted using models that also consider purely elastic host rock \cite[e.g., ][]{Mogi1958, Okada1985, Sun1969, Fialko2001}, even if evidence of inelastic deformation are visible at the Earth surface. In addition, these models are static, \textit{i.e.} they do not account for intrusion propagation, although seismological measurements evidence distributed inelastic failure of the host rock in the vicinity of propagating intrusions \cite{Roman2006, Daniels2012}.}

\textcolor{black}{Despite such geological and geophysical evidences, inelastic deformation keeps being neglected in most models of sill and laccolith emplacement. A classic argument to justify this assumption is that inelastic deformations are restricted to zones that are very small compared to the size of the modeled intrusions, and so these deformations are likely to have a negligible effect \cite[e.g.,][]{Pollard1973, Kerr1998, Bunger2011}. This assumption, however, has not been tested, so that the real effect of inelastic deformation on intrusion propagation is currently unknown. This leads to the following questions: What is the relative contribution of inelastic versus elastic deformation of the host rock during sills and laccoliths emplacement? What is the size of the inelastic zone at the tips of sills and laccoliths? To address these questions, in this paper, we develop and use a new elasto-plastic theoretical model of sill and laccolith emplacement. Here plasticity will be taken as a first, mathematically tractable, example of inelastic process. Note that due to model assumptions discussed later on, we mostly focus on the emplacement of igneous intrusions in undeformed sedimentary basins. }

The paper is structured as follows. In section 2, we build on the classic clamped elastic model of \cite{Pollard1973}, and introduce a \textcolor{black}{plastic zone} at the intrusion's tip. Unfortunately, this simple model cannot be used to uniquely determine sill growth. Therefore, in section 3, we introduce a new elasto-plastic model based on the recent model of \cite{Galland2013} and use it to predict how the plastic zone evolves as a sill grows. In section 4, we discuss the geological implications of the model.

\section{The clamped plastic model}
\label{clamped}

\subsection{Model equations}

\begin{figure}
\begin{center}
\includegraphics[width=0.8\columnwidth]{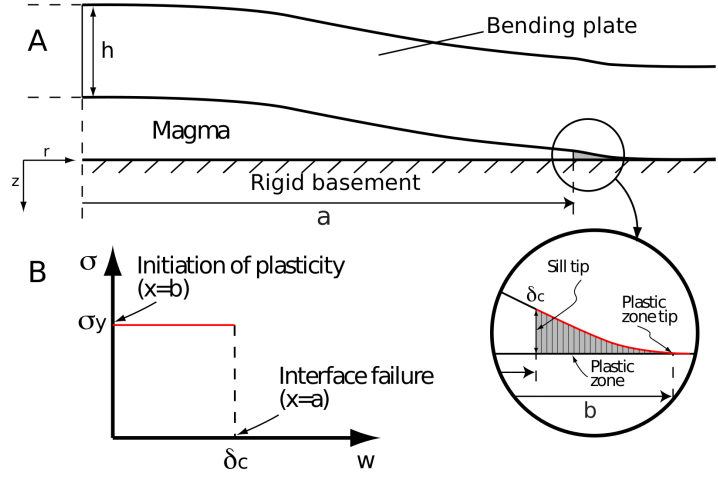}
\caption{A. Schematic drawing of the clamped plastic model. A plate of thickness $h$ is attached to a rigid foundation and is subject to the lithostatic stress $q_0=\rho g h$. An axisymmetric sill of radius $a$ applies a (possibly heterogeneous) pressure distribution $P(r)$ at the bottom of the plate and lifts it up. Between the tip of the sill ($x=a$) and the clamped region (plastic zone tip, $x=b$), a cohesive crack tip of size $b-a$ defines a \textcolor{black}{plastic} zone. The failure of the interface along which the sill propagates is defined by a critical displacement $\delta_c$. B. Schematic diagram representing the rigid-perfectly-plastic law used within the \textcolor{black}{plastic} zone illustrated in A. Plasticity is here defined by a constant stress value $\sigma_Y$, \textit{i.e.}, the yield stress of the interface between the rigid foundation and the overlying elastic plate, when the plate displacement $w$ is between $w=0$ (at $x=b$) and $w=\delta_c$ (at $x=a$).}
\label{figure2}
\end{center}
\end{figure}

We consider the following system, sketched in Fig.~\ref{figure2}: an axisymmetric flat intrusion of radius $a$ lying under a linear elastic strata of thickness $h$, Young modulus $E$, Poisson ratio $\nu$ and mass density $\rho$. We assume that the intrusion is shallow ($a/h>5$), so that the strata can be considered as a thin plate with a bending stiffness $D=\frac{Eh^3}{12(1-\nu^2)}$. Above the intrusion (radial distance $r<a$), the plate is submitted to a radial pressure profile of the form $P = P_0 - (P_0-P_a) (r/a)^n$, in which $P_0$ and $P_a$ are the pressure values at the center ($r$=0) and periphery ($r$=$a$) of the intrusion, respectively, \textcolor{black}{and $n$ is an exponent that controls the shape of the pressure field (see Fig.~2d in \cite{Galland2013})}. 

Just outside the intrusion ($r>a$), there is an \textcolor{black}{inelastic} zone in which the stress borne by the interfacial material equals its yield stress $\sigma_Y$. This is different from the purely elastic fracture assumed in the classical clamped model, where the transition between the broken and non-broken states of the interface layer is infinitely sharp. Here we define a zone of finite size that accommodates the progressive breaking process. \textcolor{black}{Field observations show that various inelastic deformation mechanisms are associated with igneous intrusion propagation: joints and micro-fractures \cite{Delaney1981}, brittle and ductile faulting \cite[e.g., ][]{Pollard1973b, Pollard1975, Schofield2012, SpacapanSubmitted}, or secondary fluidisation \cite{Schofield2012, Jackson2013}. It is challenging to account for each individual mechanism, therefore we apply a generic perfectly-plastic rheological law in the inelastic zone, subsequently referred to as plastic zone}. We define $r=b$ as the tip of the plastic zone, the length of which is thus $l_p=b-a$. Note that $b$ would be equal to $a$ ($l_p=0$) in the case of purely brittle behavior.

Outside the plastic zone, the plate is rigidly attached to the basement. At all points of the model, the strata is also submitted to the lithostatic stress $q_0=\rho g h$, with $g$ being the gravitational acceleration. In the following, we will define $\Delta P = P_0-q_0$ as the overpressure at the sill's center. Note that the basement is considered to be perfectly rigid. As sill expansion rates are much smaller than the speed of sound in the surrounding rocks and magma, we can neglect any inertial effect so that the model becomes quasi-static.

From thin plate theory, \textcolor{black}{\textit{i.e.} when the vertical displacements of the plate, $w$, remain small compared to the plate thickness $h$, }we can write the equilibrium equations of the system as:
\begin{eqnarray}
D \Delta^2 w = q_0 - P_0 +(P_0-P_a) (r/a)^n, \quad \textcolor{black}{0\le r \le a}, \label{centre}\\
D \Delta^2 w = q_0 + \sigma_Y, \quad \textcolor{black}{a \le r \le b}, \label{plastic}
\end{eqnarray}
where $\Delta^2$ is the bilaplacian operator. Note that positive displacements $w$ are defined downward, meaning that upward displacement of the plate would be negative. 

In the following sections, we will refer to $w_1$ and $w_p$ for the displacements upon the sill ($0<r<a$) and upon the plastic region ($a<r<b$), respectively. Equation (\ref{centre}), when taken in axisymmetric form with abscissa $r$, has a general solution of the form \cite[see][page 54, equation 60]{Timoshenko1959}:
\begin{equation}
w_1 = \frac{(q_0 - P_0) r^4}{64 D}+ \frac{C_1 r^2}{4} + C_2 + C_9 \ln \left(\frac{r}{b}\right) + C_{10} r^2 \ln \left(\frac{r}{b}\right) + \frac{(P_0-P_a) r^{n+4}}{D a^n (n+2)^2 (n+4)^2}.
\end{equation}
We set $C_9=C_{10}=0$ because the logarithms would lead to an unphysical displacement singularity at $r=0$. We are left with only two unknown constants, $C_1$ and $C_2$.

For $w_p$, we have to keep the contributions from the logarithms, so that:
\begin{equation}
w_p = \frac{(q_0 + \sigma_Y) r^4}{64 D}+ \frac{C_3 r^2}{4} + C_4 + C_5 \ln \left(\frac{r}{b}\right) + \textcolor{black}{C_6 r^2 \ln \left(\frac{r}{b}\right)}.
\end{equation}

We are left with the following two equations, with $C_1$ to $C_6$ being six unknown constants:
\begin{eqnarray}
w_1 = \frac{(q_0 - P_0) r^4}{64 D}+ \frac{C_1 r^2}{4} +C_2 +\frac{(P_0-P_a) r^{n+4}}{D a^n (n+2)^2 (n+4)^2}, \quad \textcolor{black}{0 \le r \le a} \label{solcenter} \\
w_p = \frac{(q_0 + \sigma_Y) r^4}{64 D}+ \frac{C_3 r^2}{4} + C_4 + C_5 \ln \left(\frac{r}{b}\right) + C_6 r^2 \ln \left(\frac{r}{b}\right), \quad \textcolor{black}{a \le r \le b}. \label{solplastic}
\end{eqnarray}

Six boundary conditions are required to uniquely determine the six unknown coefficients in Eqs.~(\ref{solcenter}) and (\ref{solplastic}). Given that the plate is rigidly attached to the basement outside the plastic zone, the displacement and the first derivative of the displacement at $r=b$ must be $0$, \textit{i.e.}:
\begin{eqnarray}
w_p(b)=0 \label{BCC5},\\
w_p'(b)=0, \label{BCC6}
\end{eqnarray}
where the prime denotes derivation with respect to $r$.

Continuity of the displacement $w$ and its three first derivatives with respect to $r$ at $r=a$ yield four boundary conditions: 
\begin{eqnarray}
w_1(a)=w_p(a) \label{BCC1},\\
w_1'(a)=w_p'(a) \label{BCC2},\\
w_1''(a)=w_p''(a) \label{BCC3},\\
w_1'''(a)=w_p'''(a) \label{BCC4}.
\end{eqnarray}

Substitution of Eqs.~(\ref{solcenter}) and (\ref{solplastic}) into Eqs.~(\ref{BCC5}) to (\ref{BCC4}) yields a linear system of six equations for the coefficients $C_1-C_6$. The system of equations is written out in full and solved in Appendix A. Note that we provide, as Supplementary Material, \textcolor{black}{both a Mathematica notebook with the analytical solutions for $C_1-C_6$ and} a Matlab code (SGHClampedPlastic.m) which calculates $C_1-C_6$ for any set of parameters ($h$, $E$, $\nu$, $\rho$, $\sigma_Y$, $P_a$, $n$, $a$, $b$ and $P_0$). \textcolor{black}{Also note that for the rest of section~\ref{clamped}, we will consider the particular case of a constant pressure distribution, $P_a=P_0$.}

\subsection{Model behavior}

\begin{figure}
\includegraphics[width=0.49\columnwidth]{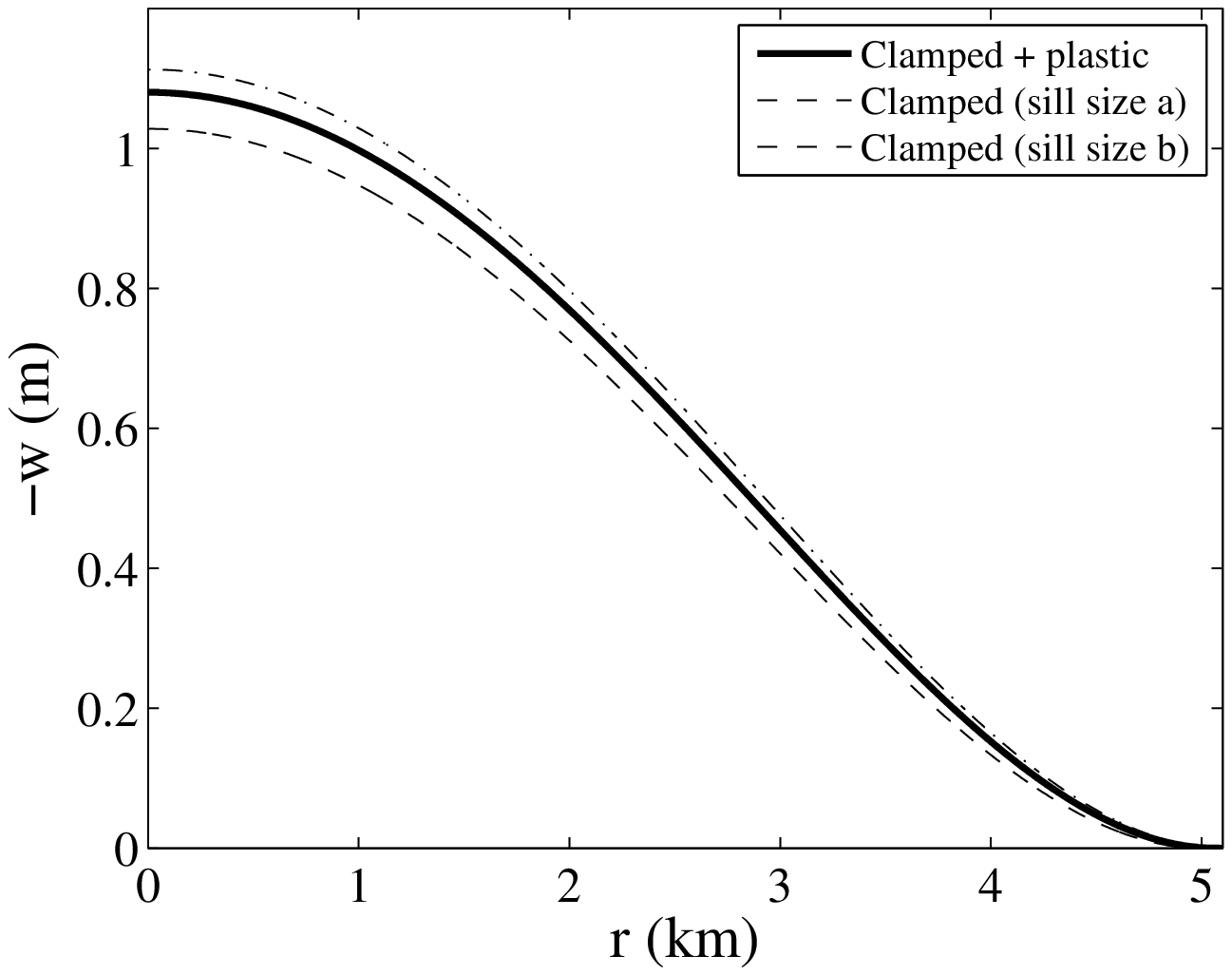}
\includegraphics[width=0.49\columnwidth]{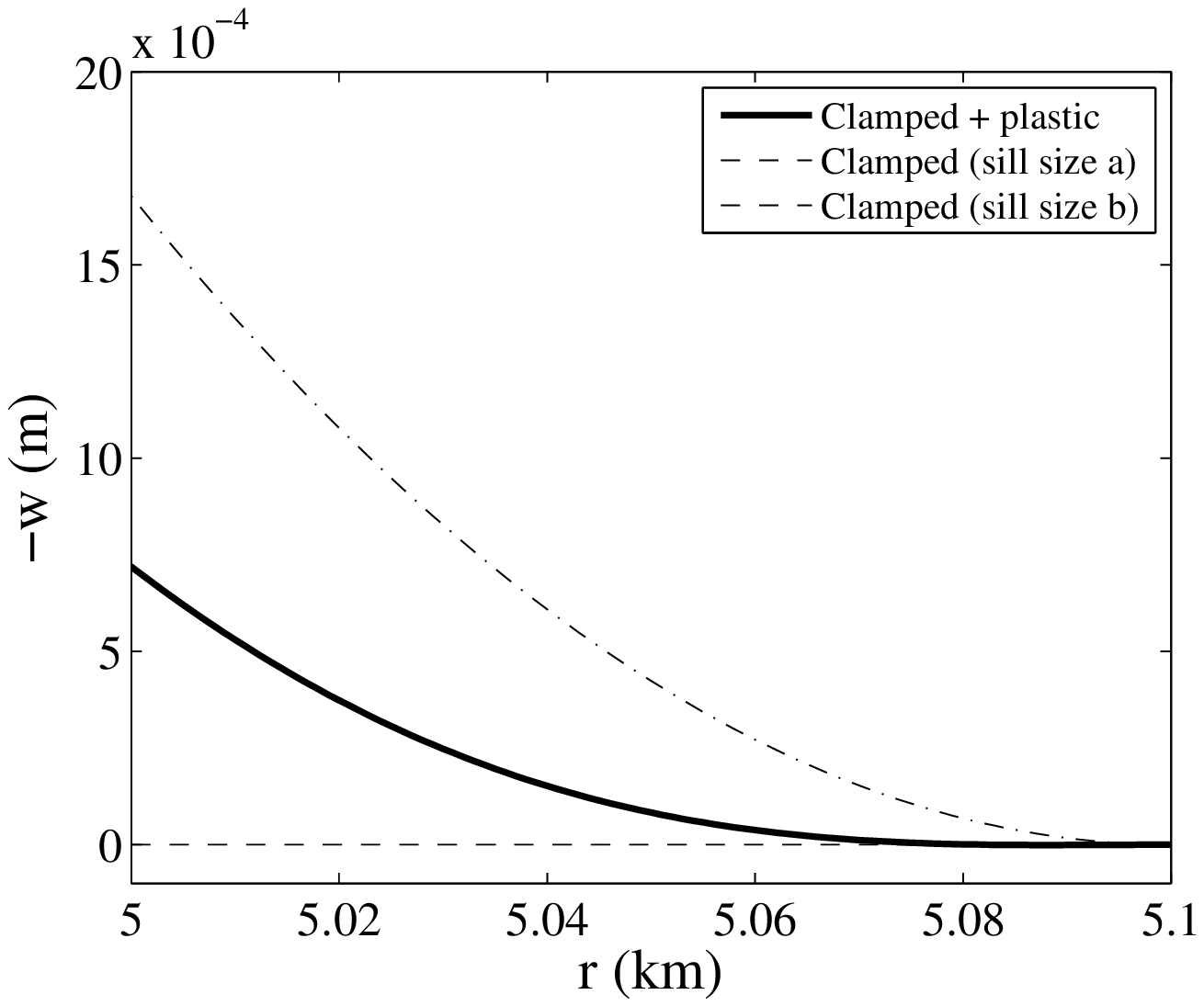}
\caption{Left: typical uplift profile $-w(r)$ for the clamped model with plasticity (solid line). Model parameters are: $h=1$ km, $E=10^{10}$ Pa, \textcolor{black}{$\nu=0.35$, $\rho=2500$ kg m$^{-3}$}, $P_0=P_a = q0 + 10^5$ Pa, $\sigma_Y=5.10^7$ Pa, $a=5$ km, $b=5.1$ km. It is compared with two profiles calculated from the \cite{Pollard1973} model ($w_{PJ}(r)=\frac{q_0-P_0}{64 D} (r^2-r_{sill}^2)^2$), with the sill radius being either $r_{sill} = a$ (dashed line) or $r_{sill} = b$ (dashed-dotted line). Right: same curves, zoomed in on the plastic zone (between $a$ and $b$).}
\label{figure3}
\end{figure}

We calculate a radial uplift profile, $-w(r)$ \textcolor{black}{(the minus sign is due to our orientation convention for $w$ and ensures that uplift is counted positively}), of \textcolor{black}{the deforming plate of thickness $h$, using} our clamped model with \textcolor{black}{plastic} zone, and compare it to the purely elastic clamped model of \cite{Pollard1973} using a set of geologically realistic parameters (Fig.~\ref{figure3}). The uplift calculated with our model is everywhere larger than that calculated with the model of \cite{Pollard1973} with the sill radius $r_{sill}=a$ (Fig.~\ref{figure3}). Conversely, the uplift calculated with our model is everywhere smaller than that calculated with the model of \cite{Pollard1973} with the sill radius $r_{sill}=b$ (Fig.~\ref{figure3}). This bracketing of our model can be readily understood by considering the uplift within the interval $a<r<b$ for the three models (Fig.~\ref{figure3}, right). In the plastic zone, the strata is allowed to deform somehow, so that the uplift is higher than for the \cite{Pollard1973} model with $r_{sill}=a$, for which the uplift vanishes by definition beyond $r=a$. The difference with the \cite{Pollard1973} model with $r_{sill}=b$ is due to the fact that the magma pressure $P_0$ pushes the strata upwards within the interval $a<r<b$, whereas, in the same interval of the plastic model, plasticity is resisting uplift.  

\begin{figure}
\begin{center}
\includegraphics[width=0.8\columnwidth]{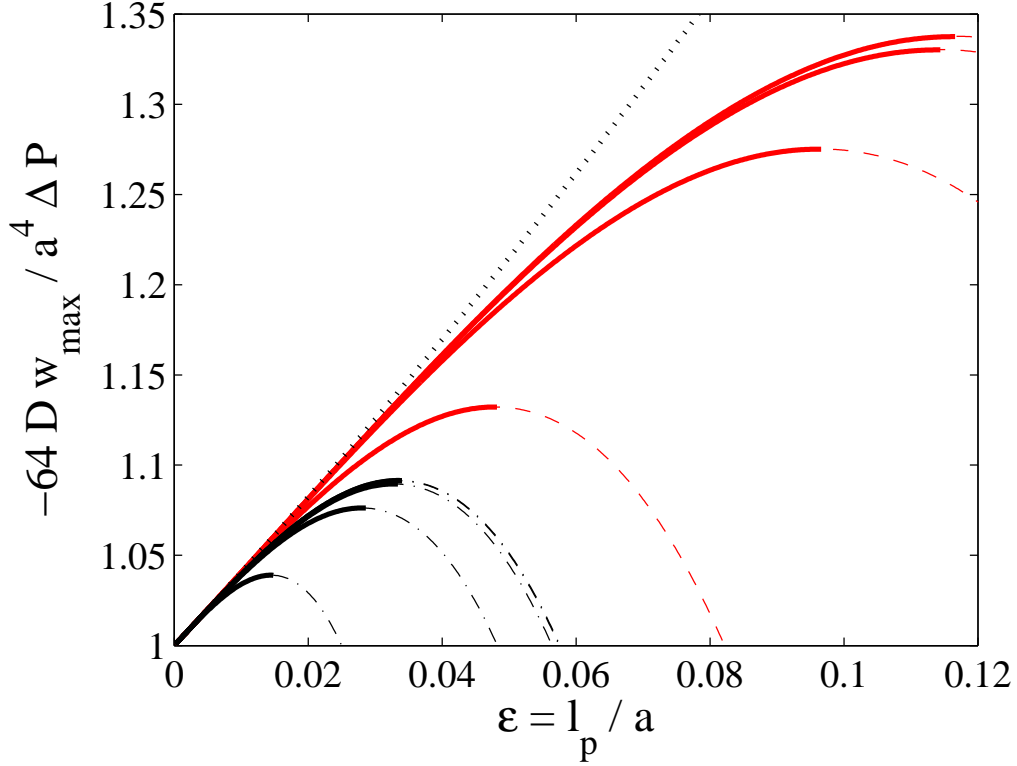}
\caption{Rescaled maximum uplift $-\frac{64 D w_{max}}{\Delta P a^4}$ as a function of the rescaled length of the plastic zone $\epsilon=\frac{l_p}{a}=\frac{b-a}{a}$. Solid lines: results of the clamped plastic model. Only values of $\epsilon$ such that the uplift is everywhere positive on $r<b$ are shown. Straight dotted line: Taylor expansions of the model result $-\frac{64 D w_{max}}{\Delta P a^4}$ for small $\epsilon$ (see Appendix \ref{AppB}, Eq.~(\ref{centredisplapprox})), truncated at second order. Dashed and dashed-dotted lines: Taylor expansions (Eq.~(\ref{centredisplapprox})), truncated at third order. Red: $\Delta P = 10^6$ Pa. Black: $\Delta P = 10^5$ Pa. For each value of $\Delta P$, four values of $\sigma_Y$ are used. From top to bottom: $\sigma_Y=10^i$ Pa, with $i$ from 5 to 8. Other model parameters are: $h=1$ km, $E=10^{10}$ Pa, \textcolor{black}{$\nu=0.35$, $\rho=2500$ kg m$^{-3}$}, $a=5$ km.}
\label{figure4}
\end{center}
\end{figure}

We want to quantify the effect of the size $l_p$ of the plastic zone, which is the unknown primary quantity of interest in our model, on the system's behavior. Following \cite{Galland2013}, we scale the maximum uplift $-w_{max}$ from our model by the maximum uplift from the clamped model $\frac{a^4 \Delta P}{64 D}$ \cite{Pollard1973}. We plot in Fig.~\ref{figure4} the results as a function of the dimensionless parameter $\epsilon=l_p/a$, which is the relative size of the plastic zone with respect to the \textcolor{black}{radius} of the sill. The advantage of this scaling is that $-\frac{64 D w_{max}}{a^4 \Delta P}=1$ when $\epsilon=0$. Figure~\ref{figure4} shows that, for small values of $\epsilon$, $-\frac{64 D w_{max}}{a^4 \Delta P}$ increases, until reaching a maximum, after which it decreases. This decrease at large values of $\epsilon$ is not physically meaningful: it corresponds to large values of $l_p$, which would induce strong downward pulling of the strata, and thus negative uplift. We found that requiring the uplift to be everywhere positive \textcolor{black}{happens to discard} the $\epsilon$ values for which the curves in Fig.~\ref{figure4} are decreasing. Therefore, we only consider the model behavior for small values of $\epsilon$. This is consistent with field observations suggesting that the sizes of plastic zones are much smaller than the \textcolor{black}{radii} of sills (\textit{i.e.} $\epsilon$ is small).

Figure~\ref{figure4} shows that the obtained rescaled curves depend on $P$, $q_0$ and $\sigma_Y$, but not on $a$, $E$ and $\nu$. These dependencies can be understood from the Taylor expansion of $-w_{max}$ for small $\epsilon$, provided in Appendix~\ref{AppB}, Eq.~(\ref{centredisplapprox}). This expansion, truncated at third order (solid lines in Fig.~\ref{figure4}), is compared to the full model (dashed and dashed-dotted lines in Fig.~\ref{figure4}). The truncated expansion seems to agree perfectly with the full model over the relevant range of $\epsilon$ values. It is interesting to note that the yield stress $\sigma_Y$ does not appear in the expansion before the third order (Equation~\ref{centredisplapprox}). As a matter of fact, the Taylor expansion of $-w_{max}$ truncated at second order appears as a straight line in Fig.~\ref{figure4} (dotted line), which shows that the third order is necessary to predict the correct shape of \textcolor{black}{the evolution of the rescaled maximum uplift as a function of $\epsilon$}. Thus, the third order term is not only required to capture the effect of $\sigma_Y$, but also the individual effects of $P_0$ and $q_0$, when they are not combined into $\Delta P$. As expected intuitively, an increase of $\sigma_Y$ or $q_0$ decreases the maximum uplift, whereas an increase of $P_0$ increases it.

\subsection{Size of the plastic zone}

Given that most theoretical models of sill and laccolith emplacement are purely elastic, none of them is able to predict the size of a plastic zone at intrusion tips. In order to derive a simple expression of the size of the plastic zone, we use the Taylor expansion of the uplift at the intrusion tip ($w_1(a)$) for small $\epsilon$ (Eq.~(\ref{tipdisplapprox})) and combine it with a classic propagation criterion\textcolor{black}{, $w_1(a)=\delta_c$,} based on a critical \textcolor{black}{vertical} displacement $\delta_c$ commonly used with cohesive zone formulations \cite[see e.g.,][]{Dugdale1960, Barenblatt1962, Chen2009}:
\begin{equation}
\textcolor{black}{\delta_c \approx \frac{a^4 \epsilon^2 (q_0 - P_0)}{16 D} + \frac{3 a^4 \epsilon^3 (q_0 - P_0)}{48 D} + \frac{a^4 \epsilon^4 \left(7 P_0 + 8 \sigma_Y + q_0 \right)}{64 D}} \label{tipdispldeltac}.
\end{equation}
This critical displacement $\delta_c$ is a material property and imposes a physical boundary condition $w_1(a)=-\delta_c$ at $r=a$, \textcolor{black}{which is valid at the onset of propagation}. Keeping only the second second order term in $\epsilon$ \textcolor{black}{in Eq.~(\ref{tipdispldeltac})}, the latter equation leads to a simple approximate expression of the dimensionless size of the plastic zone $\epsilon$ as a function of the model parameters and $\delta_c$:
\begin{equation}
\epsilon \approx  \sqrt{ \frac{16 \delta_c D}{a^4 \Delta P} } \label{sizeplasticzone}.
\end{equation}
This simple expression shows that the size $\epsilon$ of the plastic zone scales as $1/a^2$: the longer the sill, the smaller the plastic zone. This suggests that the growth of a sill is accompanied by a decrease in the size of the plastic zone. Equation (\ref{sizeplasticzone}) also highlights that $\epsilon$ scales as $1/\sqrt{\Delta P}$, meaning that the plastic zone also shrinks when the overpressure increases. Conversely, Eq.~(\ref{sizeplasticzone}) shows that $\epsilon$ scales as $\sqrt{ \delta_c}$ and $\sqrt{D}$, which suggests that the plastic zone is larger when the critical displacement for failure $\delta_c$ increases and when the overburden is very stiff and/or when the intrusion is deep.

\subsection{Ill-posedness of sill propagation}\label{sillprop1}

Equation (\ref{sizeplasticzone}) gives a simple relationship between the size of the plastic zone $\epsilon$, the propagation criterion $\delta_c$ and the variable model parameters $a$ and $\Delta P$. However, in reality, during the propagation of a sill these parameters are inter-dependent and not prescribed \textit{a priori} \cite{Murdoch2002, Galland2009, Rivalta2010, Galland2013}. Therefore, constraining the dynamics of the plastic zone during sill propagation requires a mathematical formulation to predict the coupled dynamics of $a$ and $\Delta P$ in addition to that of $\epsilon$.

The models of \cite{Murdoch2002}, \cite{Bunger2011}, \cite{Michaut2011} and \cite{Galland2013} show that the use of relevant boundary conditions is necessary to calculate the evolution of the \textcolor{black}{radius} of, and the overpressure inside, a growing sill. Typical boundary conditions used are (1) a propagation criterion, and (2) the time evolution of the volume $V$ of the sill \cite{Bunger2011, Galland2013}.

In our model with a \textcolor{black}{plastic} zone, as mentioned above, the propagation criterion is a critical displacement at the intrusion tip, \textit{i.e.}: 
 
\begin{equation}
\delta_c = w_1(a) = \frac{(q_0 - P_0) a^4}{64 D}+ \frac{C_1 a^2}{4} +C_2, \label{PropCrit1clamp}
\end{equation}
using Eq.~(\ref{tipdispl}).

\textcolor{black}{Integrating the uplift over the projected area of the sill, the} volume $V$ of the sill is easily calculated in cylindrical coordinates \cite{Galland2013}:
\begin{equation}
V = -2\pi \int_0^a  r \textcolor{black}{w_1}(r) \mathrm{d}r = -2\pi \left( \frac{(q_0 - P_0) a^6}{384 D}+ \frac{C_1 a^4}{16} +\frac{C_2 a^2}{2} \right). \label{volumesillclamp}
\end{equation}

In Eqs.~(\ref{PropCrit1clamp}) and (\ref{volumesillclamp}), $C_1$ and $C_2$ are complicated functions of $\Delta P$, $a$ and $b$, hence $V$ and $\delta_c$ are also non-trivial functions of $\Delta P$, $a$ and $b$. Thus, the mathematical problem has only two equations (Eqs.~(\ref{PropCrit1clamp}) and (\ref{volumesillclamp})) for three unknowns ($a$, $b$ and $\Delta P$), and therefore has no unique solution. Consequently, the clamped model with a \textcolor{black}{plastic} zone cannot be used to calculate the dynamics of the plastic zone during the growth of a sill, as already discussed by \cite{Kerr1998} and \cite{Galland2013}. In the following section, we demonstrate that introducing an elastic foundation, as described by \cite{Kerr1998} and \cite{Galland2013}, is sufficient to solve the dynamics of the plastic zone at the tip of a growing sill.

\section{The model with elasto-plastic foundation}
\label{elastoplastic}
\subsection{Model formulation}

\begin{figure}
\begin{center}
\includegraphics[width=0.8\columnwidth]{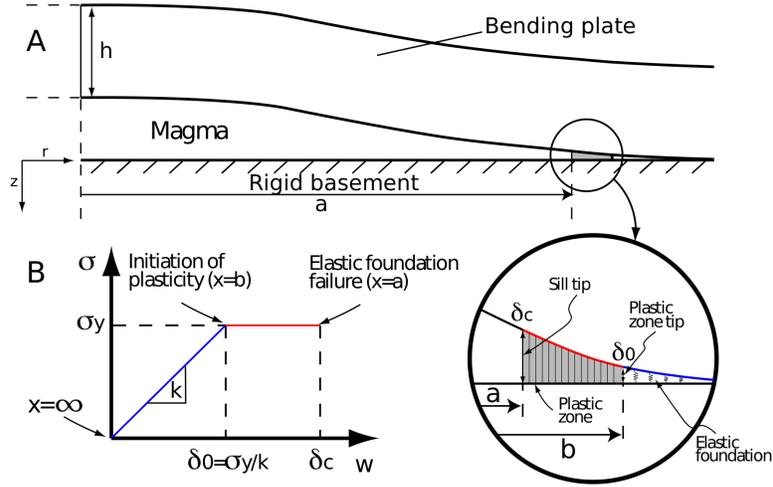}
\caption{A. Schematic drawing of the elasto-plastic model developed in this paper \cite[modified after][]{Galland2013}. A plate of thickness $h$ is attached to an elastic foundation of stiffness $k$ and is subject to the lithostatic stress $q_0$. The sill is axisymmetric with radius $a$, and a (possibly heterogeneous) pressure distribution $P(r)$ is applied at the bottom of the plate and lifts it up. Similarly to the model of \cite{Galland2013}, displacement is allowed outside the intrusion due to the elastic foundation. Here \textcolor{black}{a plastic} zone is confined between (1) the location of plasticity initiation, defined from a critical displacement $\delta_0=\frac{\sigma_Y}{k}$, and (2) the material crack tip, defined from a critical displacement $\delta_c$ dictating the failure limit of the host rock. B. Schematic diagram representing the elastic-perfectly-plastic law outside the intrusion. For small displacement $w < \delta_0$, deformation is elastic and governed by the stiffness of the elastic foundation. Displacements $\delta_0 < w < \delta_c$, define the plastic zone, in which the stress is at a constant value, \textit{i.e.} at yield stress $\sigma_Y$. For displacements $w > \delta_c$, the host rock has failed and is replaced by the over-pressurized magma.}
\label{figure5}
\end{center}
\end{figure}

We consider again the same system as described in section \ref{clamped}, but with one key difference (Fig.~\ref{figure5}): Instead of clamping the plate onto the rigid basement at $r>b$, we now assume that the plate is lying over an elastic-perfectly-plastic foundation of elastic modulus $k$ and of yield stress $\sigma_Y$. The new equilibrium equations of the system are:
\begin{eqnarray}
D \Delta^2 w = q_0 - P_0 +(P_0-P_a) (r/a)^n, \quad 0 \le r \le a, \label{centre2}\\
D \Delta^2 w = q_0 + \sigma_Y, \quad a \le r \le b, \label{plastic2}\\
D \Delta^2 w + k w = q_0, \quad r \ge b, \label{ext}
\end{eqnarray}
(plasticity of the foundation occurs between $a$ and $b$). \textcolor{black}{Again, positive displacements $w$ are defined downward, so that upward displacement of the plate is counted negatively.} 

In the case $a=b$ this model reduces to the previous model by \cite{Galland2013}. In order to check if the present model is relevant, one may use the previous model of \cite{Galland2013} to calculate $w(a)$: if $-w(a)>\frac{\sigma_Y}{k}$, then plasticity occurs, $b>a$ and the current model has to be used ; otherwise, the model of \cite{Galland2013} is sufficient.

In the following sections, we will refer to $w_1$, $w_p$ and $w_2$ for the displacements upon the sill ($0<r<a$), upon the plastic region ($a<r<b$) and outside the plastic region ($b<r$), respectively. Equation (\ref{centre2}), when taken in axisymmetric form with abscissa $r$, has a general solution of the form \cite[see][page 54, equation 60]{Timoshenko1959}:
\begin{equation}
w_1 = \frac{(q_0 - P_0) r^4}{64 D}+ \frac{C_1 r^2}{4} + C_2 + C_9 \ln \left(\frac{r}{a}\right) + C_{10} r^2 \ln \left(\frac{r}{a}\right) + \frac{(P_0-P_a) r^{n+4}}{D a^n (n+2)^2 (n+4)^2}
\end{equation}
We set $C_9=C_{10}=0$ because the logarithms would lead to a displacement singularity at $r=0$. We are left with only two unknown constants $C_1$ and $C_2$.

For $w_p$, we have to keep the contributions from the logarithms, so that:
\begin{equation}
w_p = \frac{(q_0 + \sigma_Y) r^4}{64 D}+ \frac{C_3 r^2}{4} + C_4 + C_5 \ln \left(\frac{r}{a}\right) + C_6 r^2 \ln \left(\frac{r}{a}\right).
\end{equation}
Note that the constant in the denominator within the logarithms can be chosen arbitrarily. For convenience, we use one of the length scales in the model, $a$.

The general solution of Eq.~(\ref{ext}), when the right hand side is $0$, and when taken in axisymmetric form, is provided by \cite[see][p266, equation h]{Timoshenko1959}:
\begin{equation}
w_2 = C_7 kei_0(x) + C_8 ker_0(x) + C_{11} ber_0(x) + C_{12} bei_0(x), \label{solextt}
\end{equation}
with $x=\frac{r}{l_e}$, $l_e=\sqrt[4]{\frac{D}{k}}$, and $ber_{\nu}$, $bei_{\nu}$, $ker_{\nu}$, $kei_{\nu}$ are Kelvin functions \cite{Timoshenko1959}.
We can set $C_{11}$ and $C_{12}$ to 0 because $lim_{r\rightarrow \infty}ber_0(r) = \infty$ and $lim_{r\rightarrow \infty}bei_0(r) = \infty$, which would yield unphysical infinite displacements far from the sill. Equation (\ref{ext}) also has a constant solution, $w_0=q_0/k$, which \textcolor{black}{must} be added to Eq.~(\ref{solextt}) to obtain the complete solution. Note that adding this term corresponds to the effect of the weight of the plate on the elastic foundation \cite{Galland2013}.

We are left with the following three equations, with $C_1$ to $C_8$ being eight unknown coefficients:
\begin{eqnarray}
w_1 = \frac{(q_0 - P_0) r^4}{64 D}+ \frac{C_1 r^2}{4} +C_2 +\frac{(P_0-P_a) r^{n+4}}{D a^n (n+2)^2 (n+4)^2}, \quad 0 \le r \le a, \label{solcenter2} \\
w_p = \frac{(q_0 + \sigma_Y) r^4}{64 D}+ \frac{C_3 r^2}{4} + C_4 + C_5 \ln \left(\frac{r}{a}\right) + C_6 r^2 \ln \left(\frac{r}{a}\right), \quad a \le r \le b, \label{solplastic2} \\
w_2 =  C_7 kei_0\left(\frac{r}{l_e}\right) + C_8 ker_0\left(\frac{r}{l_e}\right)  + \frac{q_0}{k}, \quad r \ge b. \label{solext}
\end{eqnarray}
To solve for the unknown coefficients, we need eight equations, which we obtain by requiring continuity of the displacement $w$ and its three first derivatives with respect to $r$ at $r=a$ and at $r=b$:
\begin{eqnarray}
w_1(a)=w_p(a) \label{BC1},\\
w_1'(a)=w_p'(a) \label{BC2},\\
w_1''(a)=w_p''(a) \label{BC3},\\
w_1'''(a)=w_p'''(a) \label{BC4}\\
w_p(b)=w_2(b) \label{BC5},\\
w_p'(b)=w_2'(b) \label{BC6},\\
w_p''(b)=w_2''(b) \label{BC7},\\
w_p'''(b)=w_2'''(b) \label{BC8}.
\end{eqnarray}
Inserting Eqs.~(\ref{solcenter2}-\ref{solext}) in (\ref{BC1}-\ref{BC8}), one obtains a set of eight linear equations for the coefficients $C_1-C_8$, which may be expressed in matrix vector form and solved by matrix inversion, as detailed in Appendix C. The analytical solutions for the coefficients are complicated, but can be found in the Mathematica notebook provided as Supplementary Material.

Replacing the values of $C_1$ to $C_8$ in Eqs.~(\ref{solcenter2}), (\ref{solplastic2}) and (\ref{solext}) provides the radial profile of vertical displacement induced by a sill for any set of system parameters ($h$, $E$, $\nu$, $\rho$, $k$, $P_a$, $\sigma_Y$ and $n$) and for any set of control parameters ($a$, $b$ and $P_0$) (Fig.~\ref{figure6}). Note that we provide as Supplementary Material a Matlab code (SGHElastoPlastic.m) which calculates $C_1-C_8$ for any set of parameters. \textcolor{black}{Also note that for the rest of section~\ref{elastoplastic}, we will consider the particular case of a constant pressure distribution, $P_a=P_0$.}

We emphasize that there are four length scales in the model: $h$, $l_e$, $a$ and $l_p=b-a$. The thickness $h$ of the elastic strata is a parameter related to the geometry of the intrusion. The elastic length $l_e=\sqrt[4]{\frac{D}{k}}$ is an intrinsic length scale of the model, which represents the lateral distance, beyond the plastic zone periphery, over which significant displacements are found \cite{Galland2013}. Note that $h$ is involved in the value of $l_e$, via $D$. 

Our model is based on Eqs.~(\ref{centre2}), (\ref{plastic2}) and (\ref{ext}), which are only valid when $a/h>>1$. In the following, we will therefore only consider values of $a$ such that $a/h>$5, with 5 being an arbitrarily chosen limit for the validity of the thin plate formulation, already used by \textit{e.g.,} \cite{Pollard1973}, \cite{Bunger2011} and \cite{Galland2013}. 

Note that before the intrusion forms, the weight of the plate already pushes down on the elastic foundation, so that there is already a homogeneous displacement $w_0=\frac{q_0}{k}$. We will consider this equilibrium state as the initial condition when the intrusion starts forming. Consequently, in order to calculate the displacement due to the intrusion, one needs to calculate the differential displacement $w_i=w-w_0=w-w(r \rightarrow \infty)$. For practical reasons, in the figures of the next sections, we plot the uplift induced by the emplacement of the intrusion, \textit{i.e.} $-w_i$ \textcolor{black}{(again, the minus sign is due to our orientation convention and ensures that uplift is counted positively)}.

\textcolor{black}{The parameter $k$ has to be interpreted as the vertical stiffness of the weak layer along which the sill propagates. An extensive discussion of its physical meaning and relationship with the mechanical properties of the weak layer, as well as the range of geologically relevant values of $k$ are provided in \cite{Galland2013}. Those values were obtained considering weak layers of minimal thickness 1m. Here, based on field observations showing thicknesses down to 10cm, we will allow for $k$ up to $10^{10}$Pa.m$^{-1}$. In practice, the smallest values of $k$ can, in the current model, lead to unrealistic negative uplift when the yield stress $\sigma_Y$ is large. As a consequence, we restricted ourselves to the range $k \in [10^7 - 10^{10}]$Pa.m$^{-1}$.}

\subsection{Model behavior}

\begin{figure}
\includegraphics[width=0.49\columnwidth]{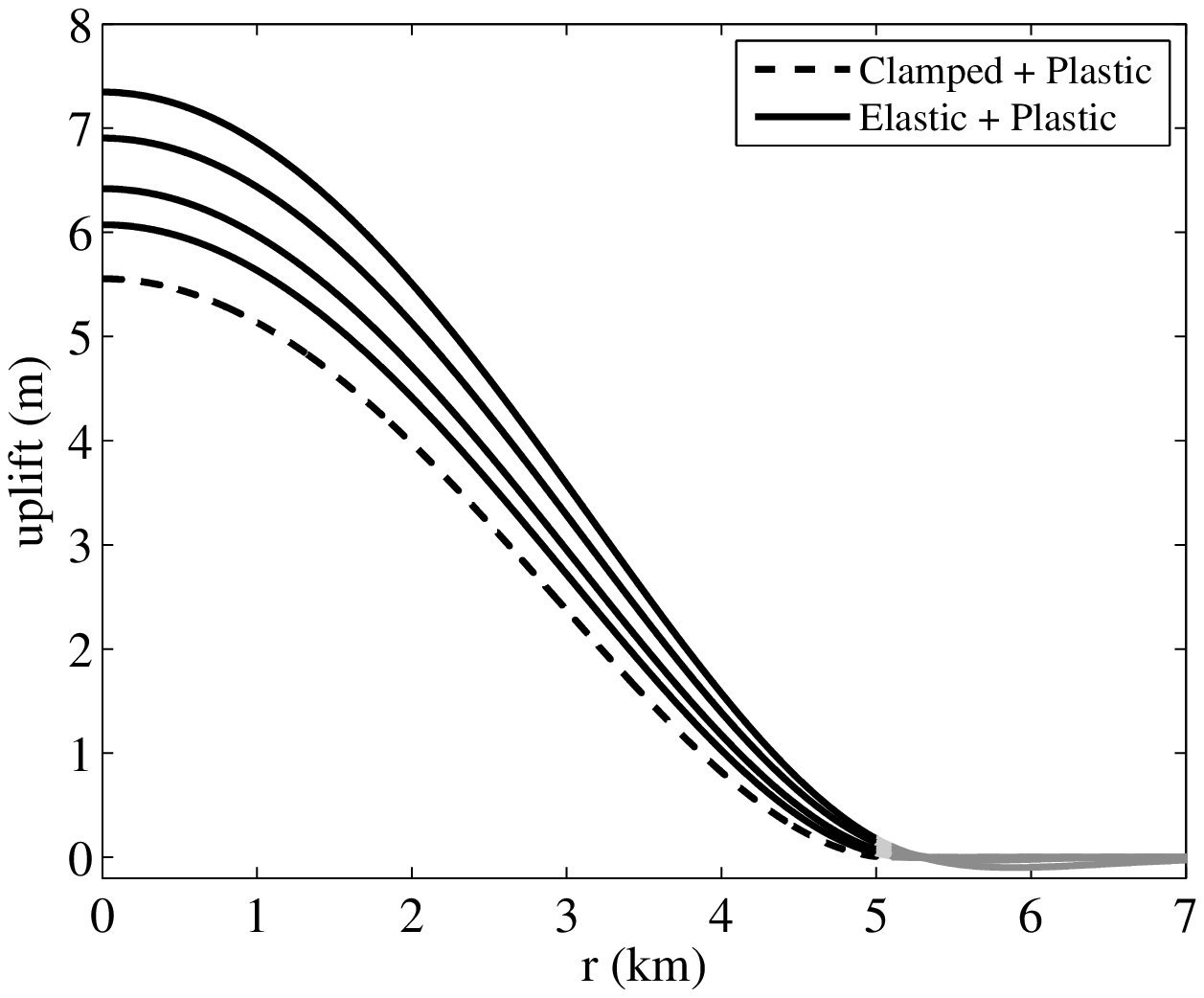}
\includegraphics[width=0.49\columnwidth]{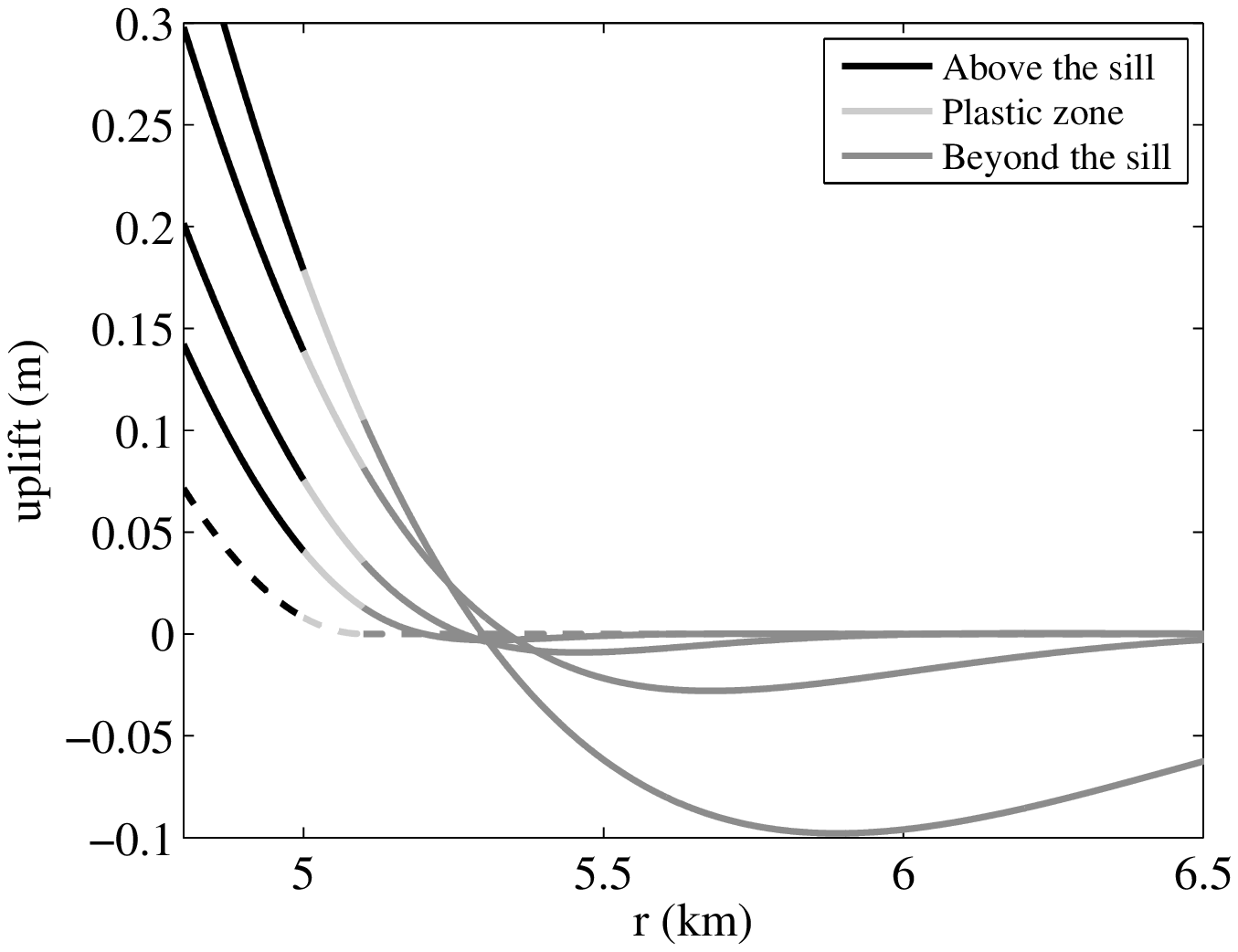}
\caption{Left: typical uplift profiles $-w_i(r)$ for the elasto-plastic model (solid lines). Model parameters are: $h=1$ km, $E=10^{10}$ Pa, \textcolor{black}{$\nu=0.35$, $\rho=2500$ kg m$^{-3}$}, $P_0=P_a = q0 + 5.10^5$ Pa, $\sigma_Y=10^6$ Pa, $a=5$ km, $b=5.1$ km, $k=10^i$ Pa m$^{-1}$ with $i$ from 7 to 10 (from top to bottom, respectively). These profiles are compared with the corresponding clamped-plastic profile, displayed with a dashed line. Right: The same curves, zoomed in at the vicinity of the plastic zone ($a<r<b$).}
\label{figure6}
\end{figure}

In this section, we investigate the behaviour of the elasto-plastic model and compare it to the clamped-plastic model described in the section~\ref{clamped}. Figure~\ref{figure6} shows, for geologically realistic parameters, typical radial uplift profiles for the elasto-plastic model. As described by \cite{Galland2013}, the uplift decreases as the stiffness of the elastic foundation increases, and the profiles converge towards the one predicted by the clamped-plastic model when the stiffness approaches infinity. The uplift outside the plastic zone is now non-zero, which is expected with an elastic foundation: similarly to the model of \cite{Galland2013}, it shows a positive uplift close to the sill's tip and a negative rebound at larger distances. Note that here the uplift at the sill's tip ($r=a$) is controlled not only by the compliance of the elastic foundation, but also by the allowed plastic deformation.

\begin{figure}
\begin{center}
\includegraphics[width=0.8\columnwidth]{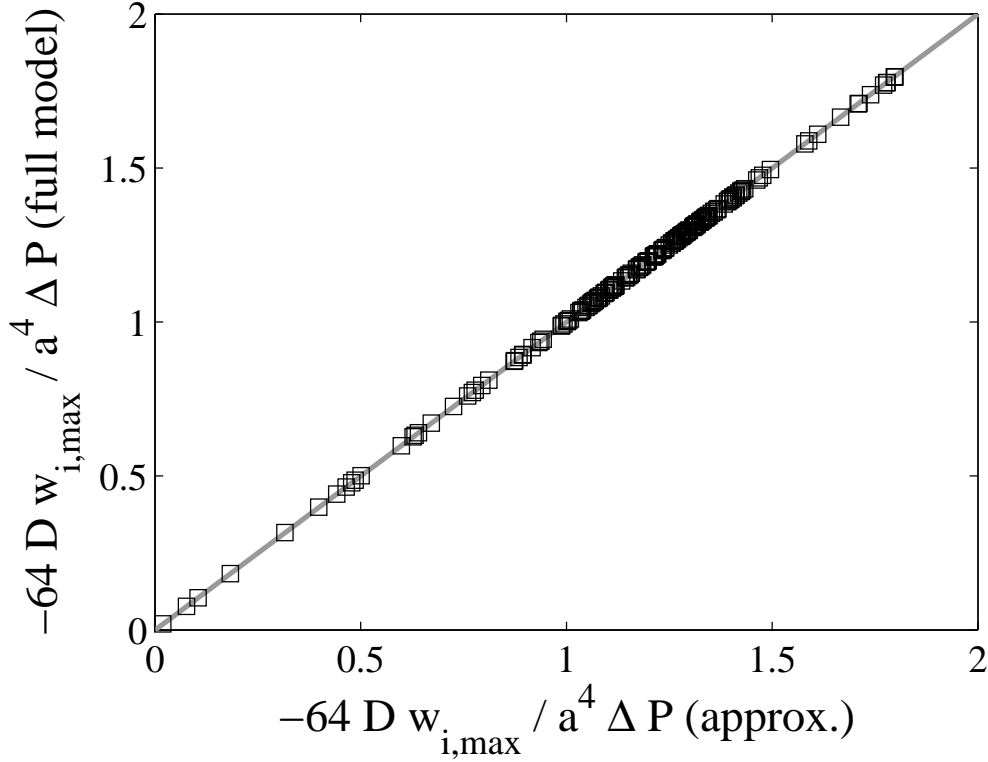}
\caption{Rescaled maximum uplift in the full elasto-plastic model, as a function of the rescaled approximate maximum uplift given by Eq.~(\ref{ApproxMaxUplift}). Squares are data points for the following sets of parameters: $h=1$ km, $E=10^{10}$ Pa, \textcolor{black}{$\nu=0.35$, $\rho=2500$ kg m$^{-3}$}, $a=5$ km, $P_0=P_a = q_0 + 10^5$ or $10^6$ Pa, $\sigma_Y=10^i$ Pa with $i=5$, $6$ or $7$, $k=10^i$ Pa m$^{-1}$ with $i$ from 7 to 10. For each set of parameters, $b$ takes 9 values such that the maximum uplift is never negative. Solid gray line: line of slope 1 passing through the origin.}
\label{figure7}
\end{center}
\end{figure}

Although the full analytic solution of the elasto-plastic model is complex, we managed to find a simple approximate analytical solution for the maximum uplift, $-w_{i,max}$, which is given in Eq.~(\ref{ApproxMaxUplift}) in Appendix~\ref{AppD}. The approximation consists in replacing the Kelvin functions in Eq.~(\ref{solextt}) by their asymptotic forms for large values of their argument $r/l_e$. Note that this approximation was previously used in \cite{Galland2013}. Figure~\ref{figure7} shows, for 216 different sets of geologically realistic parameters, that the prediction of the approximate maximum uplift captures perfectly the behavior of the full model. Equation~(\ref{ApproxMaxUplift}) can thus be used, for all practical purposes, as an excellent estimate of the maximum uplift ($-w_{i,max}$) in the model as a function of system and control parameters.

\subsection{Modeling sill propagation}

We adopt here a similar approach to that described by \cite{Galland2013} and in Section \ref{sillprop1}. Instead of treating $a$, $b$ and $\Delta P$ as model input parameters, we define three boundary conditions, in order to calculate these three quantities during the propagation of a sill. In many laboratory models \cite{Murdoch2002, Bunger2005, Galland2009, Galland2012, Galland2014} and theoretical/numerical models \cite{Murdoch2002, Malthe-Sorenssen2004, Bunger2011, Galland2013}, the growth of a sill is imposed by a constant influx rate $Q$, such that the volume of the sill at any time $t$ is known as $V(t)=Qt$. The volume of the sill, \textcolor{black}{given by $V = -2\pi \int_0^a  r w_i(r) \mathrm{d}r$}, can thus be used as a boundary condition.

Figure~\ref{figure5} highlights that in our elasto-plastic model, both sides of the plastic zone are imposed by a critical displacement. The critical displacement $\delta_0$ at the external tip of the plastic zone ($r=b$) marks the initiation of plasticity after a critical elastic displacement of the elastic foundation. The formulation of our model is such that $\delta_0$ is a direct function of the stiffness $k$ of the elastic foundation and the yield stress $\sigma_Y$, \textit{i.e.} $\delta_0=-\sigma_Y / k$. The critical displacement $\delta_c$ at the tip of the sill ($r=a$) marks the failure of the host rock. The volume boundary condition and the two critical displacement boundary conditions write:
\begin{eqnarray}
\delta_c=w_1(a)=\frac{(q_0 - P_0) a^4}{64 D}+ \frac{C_1 a^2}{4} +C_2  \label{PropCrit1},\\
\delta_0=-\sigma_Y/k=w_p(b)=\frac{(q_0 + \sigma_Y) b^4}{64 D}+ \frac{C_3 b^2}{4} + C_4 + C_5 \ln \left(\frac{b}{a}\right) + C_6 b^2 \ln \left(\frac{b}{a}\right) \label{PropCrit2},\\
V(a,b,\Delta P) = -2\pi \left( \frac{(q_0 - P_0) a^6}{384 D}+ \frac{C_1 a^4}{16} +\frac{C_2 a^2}{2} -  \frac{q_0 a^2}{2k} \right), \label{volumesill}
\end{eqnarray}
with $C_1 - C_6$ being complicated functions of $\Delta P$, $a$ and $b$. Equations (\ref{PropCrit1}), (\ref{PropCrit2}) and (\ref{volumesill}) define a system of three equations with three unknowns, $\Delta P$, $a$ and $b$, which means that, for any values of $\delta_c$, $\sigma_Y/k$ and $V$, it is possible to calculate numerically a unique set of values of $\Delta P$, $a$ and $b$. 

If we consider a growing intrusion with volume increasing linearly in time as $V(t)=Qt$, and constant propagation criteria $\delta_c$ and $\delta_0$, it is possible to calculate the evolution of $\Delta P$, $a$ and $b$ as a function of time by solving the system of Eqs.~(\ref{PropCrit1}), (\ref{PropCrit2}) and (\ref{volumesill}), similarly to the analysis of \cite{Kerr1998} and \cite{Galland2013}.

\begin{figure}
\begin{center}
\includegraphics[width=0.5\columnwidth]{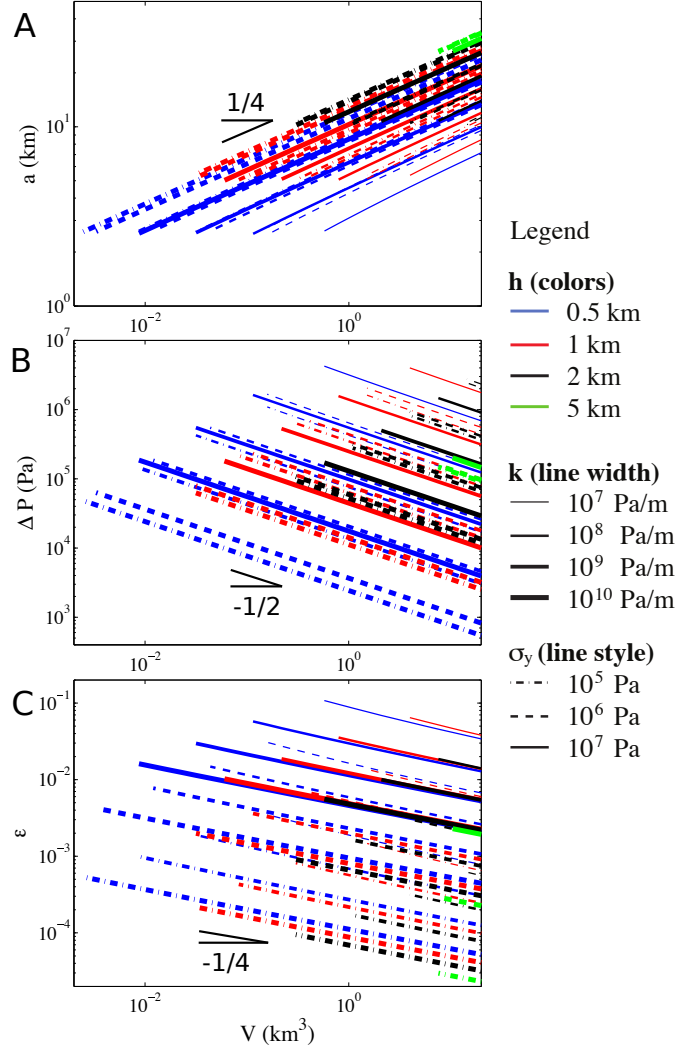}
\caption{Plots of (A) the \textcolor{black}{radius} ($a$) of the sill, (B) the magma overpressure ($\Delta P$) and (C) the size of the plastic zone ($\epsilon$), as a function of the volume $V$ of a sill during its propagation, for various values of $h$, $k$ and $\sigma_Y$. Model parameters used: $h=0.5$ km (blue), $h=1$ km (red), $h=2$ km (black), $h=5$ km (green), $k=10^7$ Pa m$^{-1}$ (thinnest line), $k=10^{8}$ Pa m$^{-1}$ (second thinnest line), $k=10^{9}$ Pa m$^{-1}$ (second thickest line), $k=10^{10}$ Pa m$^{-1}$ (thickest line), $\sigma_Y=10^5$ Pa (dotted-dashed line), $\sigma_Y=10^6$ Pa (dashed line), $\sigma_Y=10^7$ Pa (solid line).}
\label{figure8}
\end{center}
\end{figure}

Figure \ref{figure8} displays the evolutions of $\Delta P$, $a$ and $\epsilon = (b-a)/a$ during the propagation of sills for various combinations of depth $h$, foundation yield stress $\sigma_Y$ and foundation stiffness $k$. In Log-Log plots, the simulations exhibit all the same scaling. For example, we can easily show that $a \propto V^{1/4}$ (see Fig.~\ref{figure8}A) and $\Delta P \propto V^{-1/2}$ (see Fig.~\ref{figure8}B). These scaling relations are the same as those found by \cite{Murdoch2002} in the clamped elastic model and by \cite{Galland2013} in an elastic model with an elastic foundation. Such similarity likely results from the fact that in our simulations using geological values, $a/l_e \ll 1$, \textit{i.e.} the behaviour of the system is dominated by the bending plate and not by the elastic foundation \cite{Galland2013}. More interestingly, our results show that $\epsilon \propto V^{-1/4}$, \textit{i.e.} the size of the plastic zone relative to the radius of the sill decreases during the propagation of the sill. Note however that the absolute size of the plastic zone is predicted to be constant ($l_p=\epsilon a \sim V^{-1/4} V^{1/4} \sim constant$), \textit{i.e.} it does not depend on the \textcolor{black}{radius} of the propagating sill.

\begin{figure}
\begin{center}
\includegraphics[width=0.5\columnwidth]{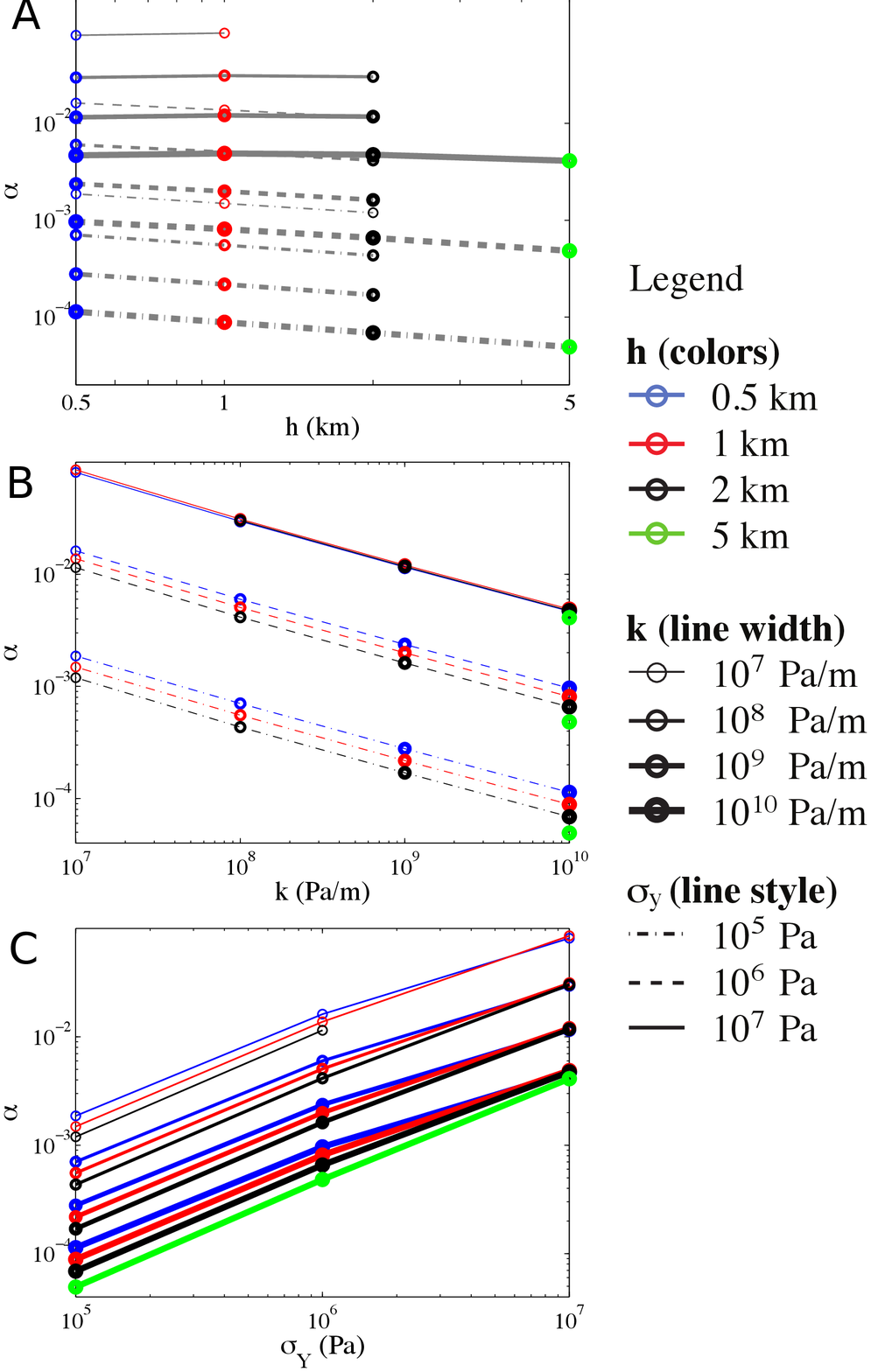}
\caption{Plots of $\alpha$, defined as $\epsilon = \alpha V^{-\frac{1}{4}}$ and calculated from data plotted in Fig.~\ref{figure8}C, as functions of the depth ($h$) (A), the stiffness of the elastic foundation ($k$) (B) and the yield stress of the elastic foundation ($\sigma_Y$) (C). Each curve of each plot considers only one variable parameter. Model parameters vary as: $h=0.5$ km (blue), $h=1$ km (red), $h=2$ km (black), $h=5$ km (green), $k=10^7$ Pa m$^{-1}$ (thinest line), $k=10^{8}$ Pa m$^{-1}$ (second thinest line), $k=10^{9}$ Pa m$^{-1}$ (second thickest line), $k=10^{10}$ Pa m$^{-1}$ (thickest line), $\sigma_Y=10^5$ Pa (dotted-dashed line), $\sigma_Y=10^6$ Pa (dashed line), $\sigma_Y=10^7$ Pa (solid line).}
\label{figure9}
\end{center}
\end{figure}

Figure~\ref{figure8}C shows that the values of $\epsilon$, for geologically relevant values of the model parameters, are all very small, with $l_p$ being typically smaller than $a/100$. This confirms that the horizontal extent of the plastic zone is confined in the close vicinity of the intrusion's tip. Figure~\ref{figure8}C also shows that $\epsilon$ greatly vary when $h$, $\sigma_Y$ and $k$ vary. Nevertheless, each curve of Fig.~\ref{figure8}C follows a function of the form $\epsilon = \alpha V^{-1/4}$. Here, comparing the values of $\alpha$ between the curves is equivalent to comparing the relative values of $\epsilon$. Figure~\ref{figure9} displays the values of $\alpha$ calculated from the data plotted in Fig.~\ref{figure8} as functions of the variable parameters $h$, $k$ and $\sigma_Y$. Each curve of each graph of Fig.~\ref{figure9} displays the dependency of $\alpha$ with respect to one variable, the two others being constant. Figure~\ref{figure9}A shows that $\alpha$ overall \textcolor{black}{slightly} decreases with increasing $h$, which shows that plastic zones are smaller for deeper sills. This result suggests that confinement at depth limits the development of plastic deformation. This conclusion, however, may lose validity for large values of $\sigma_Y$ (see Fig.~\ref{figure9}A). Figure~\ref{figure9}B shows a stronger dependency of $\alpha$ with respect to $k$: the larger $k$, the smaller $\alpha$. This is an intuitive result, which suggests that a stiff elasto-plastic foundation localizes the plastic deformation to a small \textcolor{black}{plastic} zone, and conversely weak foundations enhance the development of a broad plastic zone. Finally, Fig.~\ref{figure9}C shows that $\alpha$ increases when the yield stress increases.

\section{Interpretation and discussion}

\subsection{Model validity}

\textcolor{black}{The present model is a first attempt to include plasticity in analytic descriptions of sills and laccoliths. It is therefore oversimplified on purpose. In particular, it suffers from the same limitations as most previous elastic model~\cite{Bunger2011, Michaut2011, Galland2013, Michaut2014, Thorey2014, Michaut2016}, including linear elasticity of the deforming layer, the thin plate approximation, a single strata of homogeneous thickness, rigidity of the basement, and axisymmetric intrusions. Field observations and geophysical data show that sills and laccoliths exhibit overall sub-circular shapes in planar view, even if they are never perfectly circular. Therefore we consider our axisymmetric formulation to be relevant for addressing the main aspects of natural intrusions. In sedimentary basins, sills and laccoliths are dominantly emplaced in undeformed, flat-lying sedimentary layers. Therefore we consider that homogeneous thickness of the overburden is a relevant assumption for intrusions in sedimentary basins. In contrast, in active volcanoes, topography is rarely flat, therefore our model might have less implications for intrusions in such context. }

\textcolor{black}{A strong assumption of our model is the thin plate approximation, which implies that our model applies \textit{sensu stricto} only to shallow intrusions that fulfill the condition $a>5h$. \cite{Pollard1973} first argued that a single bending plate above sills and laccoliths is not relevant, given that their overburden is often made of stacks of sedimentary strata with different mechanical properties. They also argued that the frictional stresses between layers is most presumably much smaller than the bending stresses, so that the layers can be assumed to slide almost freely on one another. In these conditions, instead of using a single plate as thick as the intrusion's overburden, it is possible to split the overburden in many thinner plates. Doing so, the total stiffness $D$ of the layer stack is the sum of the stiffnesses $D_i$ of all layers \cite{Pollard1973}, which is always smaller than $D$.
Equivalently, the stiffness of a stack of layers of thickness $h$ has a same stiffness as a single layer with a thickness smaller than $h$. To get a sense of the implications of the layering of the bending stack of layers, let us consider a stack of $N$ layers that have approximatively the same mechanical properties and the same thickness $h_i=h/N$. The equivalent thickness of the stack, \textit{i.e.} the thickness of the single layer having the same bending stiffness, is $h_e=N^{1/3} h_i$. If we consider a $h=$1000 m thick overburden made of $N$=10 layers, this means that the equivalent thickness is about 200 m. In other words, the stack of layers is equivalent to a single layer, with a thickness $h_e$ 5 times smaller than the actual thickness $h$ of the stack. For $N$=100, $h_e$ becomes about 20 times smaller than the actual thickness. In summary, the thin plate approximation is valid when $a \gg h_e$, which considerably expands the domain of validity of our model, including sills and laccoliths with a radius $a$ possibly smaller than the depth $h$. Note that even when the layers are not identical, these conclusions remain qualitatively valid. They have been successfully applied to sills in the literature, \textit{e.g.} the Henry mountains in \cite{Pollard1973} and \cite{Koch1981} or the High Himalaya in \cite{Scaillet1995}.}

The main difference between our model and former models is the introduction of a non-linear behavior of the interfacial layer which connects the basement and the bending strata. We have implemented the two simplest plastic laws, namely rigid-perfectly-plastic (Section~\ref{clamped}) and elastic-perfectly-plastic (Section~\ref{elastoplastic}). Both of them correspond to known analytical solutions for the axisymmetric bending layer problem, within the plastic zone. Any other behavior law based on a piece-wise combination of constant and/or linear (with positive stiffness) laws as a function of vertical displacement could be used. These include elastic-plastic laws with strain-hardening \cite{Jaeger2009}, as used in \textit{e.g.}, \cite{Mazzini2009}. One would simply need to repeat the same procedure described here, \textit{i.e.}, write down the general solution for each region, apply the correct boundary conditions and solve the corresponding linear system of equations. Note that such non-linear behavior laws can also be interpreted in the framework of cohesive zone models in fracture mechanics \cite{Dugdale1960, Barenblatt1962}, as previously noted by \textit{e.g.}, \cite{Rubin1993} and \cite{Chen2009}.

We emphasize that the simple scaling of Eq.~(\ref{sizeplasticzone}) is valid only for $\epsilon \ll 1$, \textit{i.e.} when the plastic zone is small with respect to the \textcolor{black}{radius} of the intrusion. Such scaling might be lost when $\epsilon$ becomes large. Note as well that the values of $a/l_e$ in the propagation results calculated from the model with elasto-plastic foundation (Fig.~\ref{figure8}) range between 24 and 980. \cite{Galland2013} showed that for such values of $a/l_e \gg 1$, the behavior of the model with elastic foundation is dominated by the bending plate. The results and scaling calculated from the model developed in this paper (Fig.~\ref{figure8}) are thus valid under both approximations $\epsilon \ll 1$ and $a/l_e \gg 1$, which are dominantly fulfilled in natural systems. Our model might exhibit much more complex behavior if one or both approximations are not fulfilled \cite[see for example the scaling of the model of][for $a/l_e < 1$]{Galland2013}. Unravelling the full behavior of our model in a systematic manner would require extensive work, which extends beyond the scope of this paper. 

In our model, like in all sill and laccolith models using the thin plate formulation \cite[\textit{e.g.,}][]{Pollard1973, Kerr1998, Murdoch2002, Bunger2011, Galland2013}, the overlying bending plate is considered purely elastic. Recent seismic \cite{Hansen2006, Jackson2013, Magee2013} and geological \cite{Agirrezabala2015} observations, however, show that substantial parts of deformation in sills' and laccoliths' overburden is accommodated by inelastic deformations (\textit{e.g.}, compaction, fluidization, etc) in the bulk of the bending plate. Addressing such process would require further developments of our model.

\textcolor{black}{In our model, we defined a tensile propagation criterion, similarly to existing theoretical and numerical models of sill and laccolith emplacement \cite{Pollard1973, Bunger2011, Michaut2011, Galland2013, Michaut2014, Thorey2014, Michaut2016}. Note, however, that geological observations evidence some compressional deformation accommodating the propagation of sill and laccolith tips \cite[][and references therein]{Pollard1973b, Rubin1993, SpacapanSubmitted}. Accounting for this local compression in our model would require the definition of a new propagation criterion, however to our knowledge such complex mechanical propagation criterion has not been discussed in the literature.}

For sill propagation modeling purposes, we introduced a fracture criterion in terms of a critical vertical displacement ($\delta_c$) of the bending layer with respect to its unstressed state. In the literature, this critical displacement $\delta_c$ is related to the fracture energy $G_c$ of the material \cite[see \textit{e.g.},][p.31 for a table of $G_c$ values for rocks]{Scholz2002}: $G_c$ is the area under the stress-displacement curve for the interfacial material (Figs.~\ref{figure2}B and \ref{figure5}B). In our models, $\delta_c=\frac{G_c}{\sigma_Y}$ for the rigid-perfectly-plastic case used in section~\ref{clamped} and $\delta_c=\frac{G_c+\frac{\sigma_Y^2}{2k}}{\sigma_Y}$ for the elastic-perfectly-plastic case used in section~\ref{elastoplastic}.

\subsection{Geological implications}

\textcolor{black}{A first, key question that we can ask} is whether the simple, appealing scaling for $\epsilon$ in Eq.~(\ref{sizeplasticzone}) derived from the clamped model (section~\ref{clamped}) is also valid for the more advanced model with elasto-plastic foundation  (section~\ref{elastoplastic}). To address this question, we replace $a$ and $\Delta P$ in Eq.~(\ref{sizeplasticzone}) by their respective scaling $a \propto V^{1/4}$ and $\Delta P \propto V^{-1/2}$ observed during propagation within the elasto-plastic model (Figs.~\ref{figure8}A and B). This yields $\epsilon \propto V^{-1/4}$, which is indeed the propagation behaviour observed in Fig.~\ref{figure8}C. The scaling of Eq.~(\ref{sizeplasticzone}) can therefore be considered as a fundamental scaling relation for the size of the plastic zone with respect to the intrusion \textcolor{black}{radius} and magma overpressure, \textcolor{black}{ with a wide applicability. 
This result implies that the relative size of the plastic zone, $\epsilon$, decreases with increasing radius of the intrusion.
This conclusion is corroborated by the field observations of  \cite{Pollard1975}, \cite{Duffield1986}, \cite{Schofield2012, Schofield2014} and \cite{SpacapanSubmitted}, which provide evidence that plastic zones at the vicinity of the tips of small sills are sometimes as large as the sills themselves. In particular, \cite{SpacapanSubmitted} compare the extent of inelastic deformation at the tips of intrusions of distinct radii. These authors suggest that the relative size of the zone of inelastic zone decreases with the lengthening of the intrusions. Such conclusion is in very good agreement with the scaling of Eq.~(\ref{sizeplasticzone}) and our results displayed in Fig.~\ref{figure8}C. Unfortunately, since our model formulation is based on the thin plate approximation, we cannot model arbitrarily small sills and thus the very first stages of sill propagation. However, constraining the mechanics of early sill can be very helpful to constrain the dynamics of sill initiation, as demonstrated by \cite{Kavanagh2015}, who show that complex processes occur at sill inception and early growth, suggesting that plasticity might be crucial during this early stage of emplacement. Properly assessing the influence of plasticity on early sill propagation would require a different model formulation, \textit{e.g.}, the thick plate formulation \cite[see \textit{e.g.}][]{Panc1975} or Finite Element modelling. }
 
\begin{figure}
\begin{center}
\includegraphics[width=0.5\columnwidth]{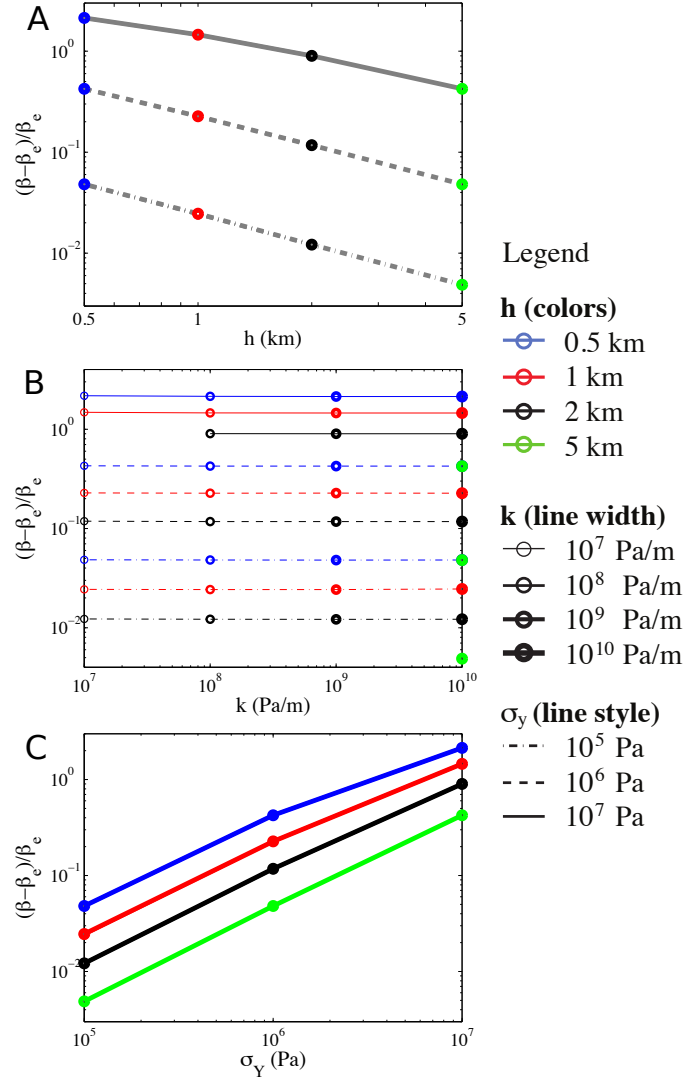}
\caption{\textcolor{black}{Relative difference of $\beta$ (defined as $\Delta P = \beta V^{-\frac{1}{2}}$ and calculated from data plotted in Fig.~\ref{figure8}B) between the elasto-plastic (this work) and the purely elastic \cite{Galland2013}) models, as a function of the depth ($h$) (A), the stiffness of the elastic foundation ($k$) (B) and the yield stress of the elastic foundation ($\sigma_Y$) (C). Each curve of each plot considers only one variable parameter. Model parameters vary as: $h=0.5$ km (blue), $h=1$ km (red), $h=2$ km (black), $h=5$ km (green), $k=10^7$ Pa m$^{-1}$ (thinest line), $k=10^{8}$ Pa m$^{-1}$ (second thinest line), $k=10^{9}$ Pa m$^{-1}$ (second thickest line), $k=10^{10}$ Pa m$^{-1}$ (thickest line), $\sigma_Y=10^5$ Pa (dotted-dashed line), $\sigma_Y=10^6$ Pa (dashed line), $\sigma_Y=10^7$ Pa (solid line).}}
\label{figure10}
\end{center}
\end{figure}

\textcolor{black}{A second, practically important question is whether the existence of inelastic processes at the tip really affect the growth dynamics of the intrusion, irrespective of the actual size of the inelastic zone. To address this question, we compared the propagation dynamics of the model with elasto-plastic foundation (this work) with that of the model with purely elastic foundation \cite{Galland2013}. We already mentioned in the description of Fig.~\ref{figure8} that the scalings of the sill's radius, $a$, and the overpressure, $\Delta P$, with the intrusion's volume, $V$ are identical for both models. The only difference is thus in the value of the prefactors of these relationships. We therefore define $\beta$ as the prefactor in $\Delta P = \beta V^{-1/2}$ for the elasto-plastic model. $\beta$ is obtained by fitting the data in Fig.~\ref{figure8}B. We define $\beta_e$ in the same way for the elastic model. Figure \ref{figure10} shows how the relative difference, $\frac{\beta-\beta_e}{\beta_e}$, between the two models, varies as a function of the model parameters $h$, $k$ and $\sigma_Y$. The differences observed range from less than 1$\%$ to as large as 200$\%$, meaning that the overpressure required to propagate the sill can be up to twice the value in the case of a purely elastic behaviour of the system. Those large differences indicate that, depending on the conditions, the effect of the (although small) plastic zone can have a major influence on the propagation dynamics of sills and laccoliths. More precisely, differences are found larger for shallower intrusions (Fig.~\ref{figure10}A) or higher values of the yield stress of the interfacial layer in which the intrusion grows (Fig.~\ref{figure10}C). Those results can be qualitatively understood by comparing the lithostatic stress, which increases with $h$, and the plastic stress $\sigma_Y$: large $h$ and/or small $\sigma_Y$ correspond to negligible plastic stress compared to the lithostatic stress. This limit precisely corresponds to the purely elastic model, and indeed the differences tend to vanish. In contrast, the stiffness of the layer has negligible effect on propagation (Fig.~\ref{figure10}B). Note that we have performed the same analysis on the prefactor $\gamma$ of the relationship $a=\gamma V^{1/4}$: the relative differences observed are found one order of magnitude smaller than those for $\beta$, for all parameters explored. The maximum observed difference of about 20$\%$ indicates that the relationship between the sill's radius and its volume is rather insensitive to the presence of plastic deformations at the intrusion's tip.}

\section{Conclusions}

In this paper we develop and use an elasto-plastic theoretical model of sill and laccolith emplacement. As in existing models, we use the formulation of a thin bending plate lying on a deformable elastic foundation. The novelty of the present study is the introduction of a cohesive plastic zone at the tip of the intrusion. The main results of our study are summarized below.

We first extended the classic clamped elastic model of \cite{Pollard1973}, and derived a fully analytic model that includes a \textcolor{black}{plastic} zone at the intrusion's tip. This model \textcolor{black}{involves} a new characteristic length: the size of the plastic zone ($l_p$). We define $\epsilon=l_p/a$, with $a$ the radius of the intrusion. The maximum uplift calculated with this model increases when $\epsilon$ increases and/or when the yield stress in the \textcolor{black}{plastic} zone ($\sigma_Y$) decreases. The model is physically meaningful only for relatively small values of $\epsilon$, but this is the range that is relevant for geological observations.

We derived a simple scaling relation for the relative size ($\epsilon$) of the \textcolor{black}{plastic} zone from the extended clamped model (Eq.~(\ref{sizeplasticzone})), which shows that $\epsilon$ scales (i) as $1/a^2$, \textit{i.e.} it is inversely proportional to the square of the intrusion's radius ($a$), and (ii) as $1/\sqrt{\Delta P}$, \textit{i.e.} it is inversely proportional to the over-pressure within the intrusion.

We demonstrate that the clamped model with \textcolor{black}{plastic} zone is not suitable for modeling the dynamics of sill propagation. We thus implemented an elasto-plastic foundation, an extension of the models of \cite{Kerr1998} and \cite{Galland2013}. \textcolor{black}{The predicted uplift is not significantly different from that predicted with the model of \cite{Pollard1973}. The most interesting outcome of the model is rather its ability to predict the evolution of the extent of the plastic zone during intrusion propagation.} Using this latter model together with a critical displacement-based propagation criterion, we show that $\epsilon$ scales with the sill's volume as $V^{-1/4}$, \textit{i.e.} the relative size of the plastic zone decreases during sill propagation. This conclusion was obtained when both approximations $\epsilon \ll 1$ and $a/l_e \gg 1$ are fulfilled. 

Our model shows that the development of a plastic zone is limited due to confinement ($\epsilon$ decreases when $h$ increases), while it is enhanced when the host rock is weak ($\epsilon$ decreases when $k$ increases).

We show that the simple scaling relation of Eq.~(\ref{sizeplasticzone}), derived from the clamped-plastic model, is also valid for the more advanced model with elasto-plastic foundation. This scaling relation is thus a fundamental characteristic of the \textcolor{black}{plastic} zone with respect to the intrusion \textcolor{black}{radius} ($a$) and magma overpressure ($\Delta P$).

\textcolor{black}{All in all, our novel elasto-plastic model highlights that although the inelastic zone is probably negligibly small for the large, shallow sills considered here, it can have a significant effect on their propagation dynamics.
We suggest that an interesting follow-up of this study would be to extend theoretical models beyond the thin plate approximation to also unravel the dynamics of early sill emplacement.}

\textit{This study was supported by Physics of Geological Processes (PGP). J.S. acknowledges support from the People Programme (Marie Curie Actions) of the European Union's 7th Framework Programme (FP7/2007-2013) under Research Executive Agency Grant Agreement 303871.}


%
\appendix
%
%

\section{Clamped model}\label{AppA}

Here we rewrite Eqs.~(\ref{BCC5}) to (\ref{BCC4}) for the clamped-plastic model, combine them in matrix form and provide the analytical solution for the six coefficients $C_1$ to $C_6$.

\textcolor{black}{Using the expression of $w_p$ given in Eq.~(\ref{solplastic}) and taken in $r=b$, Eq.~(\ref{BCC5}) can be rewritten as:}
\begin{eqnarray}
\frac{(q_0 + \sigma_Y) b^4}{64 D}+ \frac{C_3 b^2}{4} + C_4 = 0.
\end{eqnarray}

\textcolor{black}{Using the derivative of Eq.~(\ref{solplastic}) taken in $r=b$, Eq.~(\ref{BCC6}) can be rewritten as:}
\begin{eqnarray}
\frac{(q_0 + \sigma_Y) b^3}{16 D}+ \frac{C_3 b}{2} + \frac{C_5}{b} + C_6 b  = 0.
\end{eqnarray}

\textcolor{black}{Using the expressions of $w_1$ and $w_p$ given in Eqs.~(\ref{solcenter}) and (\ref{solplastic}) and taken in $r=a$, Eq.~(\ref{BCC1}) can be rewritten as:}
\begin{eqnarray}
\frac{(q_0 - P_0) a^4}{64 D}+ \frac{C_1 a^2}{4} +C_2 +\frac{(P_0-P_a) a^4}{D (n+2)^2 (n+4)^2} = \nonumber \\ 
\frac{(q_0 + \sigma_Y) a^4}{64 D}+ \frac{C_3 a^2}{4} +C_4 + C_5 \ln \left(\frac{a}{b}\right)+ C_6 a^2 \ln \left(\frac{a}{b}\right).
\end{eqnarray}

\textcolor{black}{Using the first derivatives of Eqs.~(\ref{solcenter}) and (\ref{solplastic}) taken in $r=a$, Eq.~(\ref{BCC2}) can be rewritten as:}
\begin{eqnarray}
\frac{(q_0 - P_0) a^3}{16 D}+ \frac{C_1 a}{2} + \frac{(P_0-P_a) a^3}{D (n+2)^2 (n+4)} = \nonumber \\ 
\frac{(q_0 + \sigma_Y) a^3}{16 D}+ \frac{C_3 a}{2} + \frac{C_5}{a} + C_6 a \left(1+2 \ln \left(\frac{a}{b}\right) \right).
\end{eqnarray}

\textcolor{black}{Using the second derivatives of Eqs.~(\ref{solcenter}) and (\ref{solplastic}) taken in $r=a$, Eq.~(\ref{BCC3}) can be rewritten as:}
\begin{eqnarray}
\frac{3(q_0 - P_0) a^2}{16 D}+ \frac{C_1}{2} + \frac{(P_0-P_a) a^2 (n+3)}{D (n+2)^2 (n+4)} =\nonumber \\ 
\frac{3(q_0 + \sigma_Y) a^2}{16 D}+ \frac{C_3}{2} - \frac{C_5}{a^2} + C_6 \left(3+2 \ln \left(\frac{a}{b}\right) \right).
\end{eqnarray}

\textcolor{black}{Using the third derivatives of Eqs.~(\ref{solcenter}) and (\ref{solplastic}) taken in $r=a$, Eq.~(\ref{BCC4}) can be rewritten as:}
\begin{eqnarray}
\frac{3(q_0 - P_0) a}{8 D} + \frac{(P_0-P_a) a (n+3)}{D (n+2) (n+4)} =\nonumber \\
\frac{3(q_0 + \sigma_Y) a}{8 D} + \frac{2 C_5}{a^3} + \frac{2 C_6}{a}.
\end{eqnarray}

These equations constitute a system of six coupled linear equations, which can be written matricially as :
\begin{equation}
A \dot C=B, \label{solmatrix}
\end{equation}
with $A$=\\
\begin{math}
\begin{pmatrix}
0 & 0 & \frac{b^2}{4} & 1 & 0 & 0  \\
0 & 0 & \frac{b}{2} & 0 & \frac{1}{b} & b  \\
\frac{a^2}{4} & 1 & -\frac{a^2}{4} & -1 & -\ln \left(\frac{a}{b}\right)  & -a^2 \ln \left(\frac{a}{b}\right) \\
\frac{a}{2} & 0 & -\frac{a}{2} & 0 & -\frac{1}{a} & -a \left(1 + 2 \ln \left(\frac{a}{b}\right) \right) \\
\frac{1}{2} & 0 & -\frac{1}{2} & 0 & \frac{1}{a^2} & -  \left(3 + 2 \ln \left(\frac{a}{b}\right) \right) \\
0 & 0 & 0 & 0 & -\frac{2}{a^3} & -\frac{2}{a}

\end{pmatrix}
\end{math}, 
\\ 
\\ 
\\ 
\begin{math}
B=
\left(
\begin{array}{clrr}
- \frac{(q_0 + \sigma_Y) b^4}{64 D}\\
- \frac{(q_0 + \sigma_Y) b^3}{16 D}\\
\left[\frac{(P_0 + \sigma_Y)}{64} - \frac{(P_0-P_a)}{(n+2)^2 (n+4)^2}\right] \frac{a^4}{D}\\
\left[\frac{(P_0 + \sigma_Y)}{16} - \frac{(P_0-P_a)}{(n+2)^2 (n+4)}\right] \frac{a^3}{D}\\
\left[\frac{3(P_0 + \sigma_Y)}{16} - \frac{(P_0-P_a)(n+3)}{(n+2)^2 (n+4)}\right] \frac{a^2}{D}\\
\left[\frac{3(P_0 + \sigma_Y)}{8} - \frac{(P_0-P_a)(n+3)}{(n+2) (n+4)}\right] \frac{a}{D}
\end{array}
\right)
\end{math}
and 
\begin{math}
C=
\left(
\begin{array}{clrr}
C_1 \\
C_2 \\
C_3 \\
C_4 \\
C_5 \\
C_6
\end{array}
\right)
\end{math}
\\

The solution vector $C$ has an analytic solution which is given in the Mathematica notebook provided as Supplementary Material. When considering the pressure distribution as constant ($P_0=P_a$), the following simplified expressions for the coefficients $C_1-C_6$ can be obtained: 

\begin{eqnarray}
C_1=\frac{\left( \sigma_Y + P_0 \right) a^2 \left(a^2 - 4 b^2 \ln \left( \frac{a}{b} \right) \right) - \left( q_0 + \sigma_Y \right) b^4}{8 b^2 D}  \label{CC1},\\
C_2=\frac{\left( \sigma_Y + P_0 \right) \left[3a^4 - 4 a^2 b^2 - 4 a^4 \ln \left( \frac{a}{b} \right) \right] + \left( q_0 + \sigma_Y \right) b^4}{64 D} \label{CC2},\\
C_3=\frac{\left( \sigma_Y + P_0 \right) a^2 \left(a^2 + 2 b^2 \right) - \left( q_0 + \sigma_Y \right) b^4}{8 b^2 D} \label{CC3},\\
C_4= \frac{- 2 \left( \sigma_Y + P_0 \right) a^2 \left(a^2 + 2 b^2 \right) + \left( q_0 + \sigma_Y \right) b^4}{64 D}  \label{CC4},\\
C_5=-\frac{\left( \sigma_Y + P_0 \right) a^4}{16D} \label{CC5},\\
C_6=-\frac{\left( \sigma_Y + P_0 \right) a^2}{8D} \label{CC6}.
\end{eqnarray}

These solutions allow us to obtain, for any set of system parameters ($h$, $E$, $\nu$, $\rho$, $\sigma_Y$) and for any control parameters ($a$, $b$ and $P_0$), the analytical expression of the radial profile of vertical displacement $w(r)$ (see \textit{e.g.}, Fig.~\ref{figure3}).

We provide as Supplementary Material a Matlab code (SGHClampedPlastic.m) which calculates $C_1-C_6$ for any set of parameters ($h$, $E$, $\nu$, $\rho$, $\sigma_Y$, $P_a$, $n$, $a$, $b$ and $P_0$). We also provide the analytic expressions in a Mathematica notebook.

\section{Taylor expansion of the clamped model}\label{AppB}

We \textcolor{black}{obtain} the series expansion of $C_1$ and $C_2$ with respect to $\epsilon$, for small $\epsilon$, by replacing the expression of $b=a\left(1 + \epsilon \right)$ in Eqs.~(\ref{CC1}) and (\ref{CC2}) \textcolor{black}{and by combining the terms with the same power of $\epsilon$}:

\begin{eqnarray}
C_1=\frac{a^2\left(P_0 - q_0 \right)}{8 D} + \frac{a^2 \epsilon \left(P_0 - q_0 \right)}{4 D} + \frac{a^2 \epsilon^2 \left(P_0 - q_0 \right)}{8 D} - \frac{a^2 \epsilon^3 \left(P_0 + \sigma_Y \right)}{3 D}+ \frac{a^2 \epsilon^4 \left(P_0 + \sigma_Y \right)}{2 D} + O(\epsilon^5)\label{SeriesExpC1},\\
C_2=\frac{a^4\left(q_0 - P_0 \right)}{64 D} + \frac{a^4 \epsilon \left(q_0 - P_0 \right)}{16 D} + \frac{3 a^4 \epsilon^2 \left(q_0 - P_0 \right)}{32 D} + \frac{a^4 \epsilon^3 \left(P_0 + 3 q_0 + 4 \sigma_Y \right)}{48 D} + \frac{a^4 \epsilon^4 \left( q_0 - P_0 \right)}{64 D}+ O(\epsilon^5) \label{SeriesExpC2}.
\end{eqnarray}

\textcolor{black}{Setting $r=0$ in} Eq.~(\ref{solcenter}) provides a straightforward expression of the maximum displacement $w_{max}$:

\begin{equation}
w_1(0) = w_{max} = C_2 \label{centerdispl}.
\end{equation}

Combining Eq.~(\ref{centerdispl}) with Eq.~(\ref{SeriesExpC2}) leads to an approximate expression of the maximum displacement $w_{max}$ as a function of the model parameters and $\epsilon$:

\begin{equation}
\textcolor{black}{w_{max} \approx \frac{a^4\left(q_0 - P_0 \right)}{64 D} + \frac{a^4 \epsilon \left(q_0 - P_0 \right)}{16 D} + \frac{3 a^4 \epsilon^2 \left(q_0 - P_0 \right)}{32 D} + \frac{a^4 \epsilon^3 \left(P_0 + 3 q_0 + 4 \sigma_Y \right)}{48 D} + \frac{a^4 \epsilon^4 \left( q_0 - P_0 \right)}{64 D}} \label{centredisplapprox},
\end{equation}
\textcolor{black}{which, in dimensionless form reads $-\frac{64 D w_{max}}{\Delta P a^4} \approx 1 + 4 \epsilon + 6 \epsilon^2 - \frac{4}{3}\left[1 + \frac{4 (q_0 + \sigma_Y)}{\Delta P}\right] \epsilon^3 + \epsilon^4$. This expression captures the behaviours shown in Fig.~\ref{figure4}.}

Similarly, \textcolor{black}{setting $r=a$ in} Eq.~(\ref{solcenter}) provides a straightforward expression of the displacement at the intrusion tip ($r=a$):

\begin{equation}
w_1(a) = \frac{(q_0 - P_0) a^4}{64 D}+ \frac{C_1 a^2}{4} +C_2 \label{tipdispl}.
\end{equation}

Using the expressions of $C_1$ and $C_2$ from Eqs.~(\ref{SeriesExpC1}) and (\ref{SeriesExpC2}) in Eq.~(\ref{tipdispl}), we derive an approximate expression of the displacement at the tip of the intrusion ($r=a$):

\begin{equation}
\textcolor{black}{w_1(a) \approx \frac{a^4 \epsilon^2 (q_0 - P_0)}{16 D} + \frac{3 a^4 \epsilon^3 (q_0 - P_0)}{48 D} + \frac{a^4 \epsilon^4 \left(7 P_0 + 8 \sigma_Y + q_0 \right)}{64 D}} \label{tipdisplapprox}.
\end{equation}

Note that the effect of the yield stress $\sigma_Y$ on $w_1(a)$ appears only at the fourth order of $\epsilon$.

\section{Elasto-plastic model}\label{AppC}

Here we rewrite Eqs.~(\ref{BC1}) to (\ref{BC8}) for the elasto-plastic model, combine them in matrix form and provide the analytical solution for the eight coefficients $C_1$ to $C_8$.

\textcolor{black}{Using the expressions of $w_1$ and $w_p$ in Eqs.~(\ref{solcenter2}) and (\ref{solplastic2}) and taken in $r=a$, Eq.~(\ref{BC1}) can be rewritten as:}
\begin{eqnarray}
\frac{(q_0 - P_0) a^4}{64 D}+ \frac{C_1 a^2}{4} +C_2 +\frac{(P_0-P_a) a^4}{D (n+2)^2 (n+4)^2} = \nonumber \\ 
\frac{(q_0 + \sigma_Y) a^4}{64 D}+ \frac{C_3 a^2}{4} +C_4 
\end{eqnarray}

\textcolor{black}{Using the first derivatives of Eqs.~(\ref{solcenter2}) and (\ref{solplastic2}) taken in $r=a$, Eq.~(\ref{BC2}) can be rewritten as:}
\begin{eqnarray}
\frac{(q_0 - P_0) a^3}{16 D}+ \frac{C_1 a}{2} + \frac{(P_0-P_a) a^3}{D (n+2)^2 (n+4)} = \nonumber \\ 
\frac{(q_0 + \sigma_Y) a^3}{16 D}+ \frac{C_3 a}{2} + \frac{C_5}{a} + C_6 a 
\end{eqnarray}

\textcolor{black}{Using the second derivatives of Eqs.~(\ref{solcenter2}) and (\ref{solplastic2}) taken in $r=a$, Eq.~(\ref{BC3}) can be rewritten as:}
\begin{eqnarray}
\frac{3(q_0 - P_0) a^2}{16 D}+ \frac{C_1}{2} + \frac{(P_0-P_a) a^2 (n+3)}{D (n+2)^2 (n+4)} =\nonumber \\ 
\frac{3(q_0 + \sigma_Y) a^2}{16 D}+ \frac{C_3}{2} - \frac{C_5}{a^2} + 3 C_6
\end{eqnarray}

\textcolor{black}{Using the third derivatives of Eqs.~(\ref{solcenter2}) and (\ref{solplastic2}) taken in $r=a$, Eq.~(\ref{BC4}) can be rewritten as:}
\begin{eqnarray}
\frac{3(q_0 - P_0) a}{8 D} + \frac{(P_0-P_a) a (n+3)}{D (n+2) (n+4)} =\nonumber \\
\frac{3(q_0 + \sigma_Y) a}{8 D} + \frac{2 C_5}{a^3} + \frac{2 C_6}{a}
\end{eqnarray}

\textcolor{black}{Using the expressions of $w_p$ and $w_2$ in Eqs.~(\ref{solplastic2}) and (\ref{solext}) and taken in $r=b$, Eq.~(\ref{BC5}) can be rewritten as:}
\begin{eqnarray}
\frac{(q_0 + \sigma_Y) b^4}{64 D}+ \frac{C_3 b^2}{4} + C_4 + C_5 \ln \left(\frac{b}{a}\right) + C_6 b^2 \ln \left(\frac{b}{a}\right)= \nonumber \\ 
C_7 kei_0\left(\frac{b}{l_e}\right) + C_8 ker_0\left(\frac{b}{l_e}\right) + \frac{q_0}{k}
\end{eqnarray}

\textcolor{black}{Using the first derivatives of Eqs.~(\ref{solplastic2}) and (\ref{solext}) taken in $r=b$, Eq.~(\ref{BC6}) can be rewritten as:}
\begin{eqnarray}
\frac{(q_0 + \sigma_Y) b^3}{16 D}+ \frac{C_3 b}{2} + \frac{C_5}{b} + C_6 b \left( 2 \ln \left(\frac{b}{a}\right)+1 \right)  = \nonumber \\ 
\frac{C_7}{\sqrt{2}l_e} \left[kei_1\left(\frac{b}{l_e}\right)-ker_1\left(\frac{b}{l_e}\right)\right] + \frac{C_8}{\sqrt{2}l_e} \left[kei_1\left(\frac{b}{l_e}\right)+ker_1\left(\frac{b}{l_e}\right)\right]
\end{eqnarray}

\textcolor{black}{Using the second derivatives of Eqs.~(\ref{solplastic2}) and (\ref{solext}) taken in $r=b$, Eq.~(\ref{BC7}) can be rewritten as:}
\begin{eqnarray}
\frac{3(q_0 + \sigma_Y) b^2}{16 D}+ \frac{C_3}{2} - \frac{C_5}{b^2} + C_6 \left( 2 \ln \left(\frac{b}{a}\right) +3 \right)=\nonumber \\ 
\frac{C_7}{2l_e^2} \left[ker_0\left(\frac{b}{l_e}\right)-ker_2\left(\frac{b}{l_e}\right)\right] + \frac{C_8}{2l_e^2} \left[kei_2\left(\frac{b}{l_e}\right)-kei_0\left(\frac{b}{l_e}\right)\right]
\end{eqnarray}

\textcolor{black}{Using the third derivatives of Eqs.~(\ref{solplastic2}) and (\ref{solext}) taken in $r=b$, Eq.~(\ref{BC8}) can be rewritten as:}
\begin{eqnarray}
\frac{3(q_0 + \sigma_Y) b}{8 D} + \frac{2 C_5}{b^3} + \frac{2 C_6}{b} =\nonumber \\ 
\frac{C_7}{4\sqrt{2}l_e^3} \left[3 ker_1\left(\frac{b}{l_e}\right)-ker_3\left(\frac{b}{l_e}\right)+3 kei_1\left(\frac{b}{l_e}\right)-kei_3\left(\frac{b}{l_e}\right)\right]\nonumber \\ 
+ \frac{C_8}{4\sqrt{2}l_e^3} \left[3 ker_1\left(\frac{b}{l_e}\right)-ker_3\left(\frac{b}{l_e}\right)-3 kei_1\left(\frac{b}{l_e}\right)+kei_3\left(\frac{b}{l_e}\right)\right]
\end{eqnarray}

These equations constitute a system of eight coupled linear equations, which can be written matricially as :
\begin{equation}
A.C=B, \label{solmatrix}
\end{equation}
with $A$=\\
\begin{math}
\begin{pmatrix}
\frac{a^2}{4} & 1 & -\frac{a^2}{4} & -1 & 0 & 0 & 0 & 0 \\
\frac{a}{2} & 0 & -\frac{a}{2} & 0 & -\frac{1}{a} & -a & 0 & 0 \\
\frac{1}{2} & 0 & -\frac{1}{2} & 0 & \frac{1}{a^2} & -3 & 0 & 0 \\
0 & 0 & 0 & 0 & -\frac{2}{a^3} & -\frac{2}{a} & 0 & 0 \\
0 & 0 & \frac{b^2}{4} & 1 & \ln \left(\frac{b}{a}\right) & b^2 \ln \left(\frac{b}{a}\right) & -kei_0(\frac{b}{l_e}) & -ker_0(\frac{b}{l_e}) \\
0 & 0 & \frac{b}{2} & 0 & \frac{1}{b} & b \left( 2 \ln \left(\frac{b}{a}\right) + 1 \right) & -\frac{kei_1(\frac{b}{l_e})-ker_1(\frac{b}{l_e})}{\sqrt{2}l_e} & -\frac{kei_1(\frac{b}{l_e})+ker_1(\frac{b}{l_e})}{\sqrt{2}l_e} \\
0 & 0 & \frac{1}{2} & 0 & -\frac{1}{b^2} & 2 \ln \left(\frac{b}{a}\right) + 3 & -\frac{ker_0(\frac{b}{l_e})-ker_2(\frac{b}{l_e})}{2l_e^2} & -\frac{kei_2(\frac{b}{l_e})-kei_0(\frac{b}{l_e})}{2l_e^2} \\
0 & 0 & 0 & 0 & \frac{2}{b^3} & \frac{2}{b} & -\frac{3 ker_1(\frac{b}{l_e})-ker_3(\frac{b}{l_e})+3 kei_1(\frac{b}{l_e})-kei_3(\frac{b}{l_e})}{4\sqrt{2}l_e^3} & -\frac{3 ker_1(\frac{b}{l_e})-ker_3(\frac{b}{l_e})-3 kei_1(\frac{b}{l_e})+kei_3(\frac{b}{l_e})}{4\sqrt{2}l_e^3} 

\end{pmatrix}
\end{math}, 
\\ 
\\ 
\\ 
\begin{math}
B=
\left(
\begin{array}{clrr}
\left[\frac{(P_0 + \sigma_Y)}{64} - \frac{(P_0-P_a)}{(n+2)^2 (n+4)^2}\right] \frac{a^4}{D}\\
\left[\frac{(P_0 + \sigma_Y)}{16} - \frac{(P_0-P_a)}{(n+2)^2 (n+4)}\right] \frac{a^3}{D}\\
\left[\frac{3(P_0 + \sigma_Y)}{16} - \frac{(P_0-P_a)(n+3)}{(n+2)^2 (n+4)}\right] \frac{a^2}{D}\\
\left[\frac{3(P_0 + \sigma_Y)}{8} - \frac{(P_0-P_a)(n+3)}{(n+2) (n+4)}\right] \frac{a}{D}\\
\frac{q_0}{k} - \frac{(q_0 + \sigma_Y) b^4}{64 D}\\
- \frac{(q_0 + \sigma_Y) b^3}{16 D}\\
- \frac{3(q_0 + \sigma_Y) b^2}{16 D}\\
- \frac{3(q_0 + \sigma_Y) b}{8 D}
\end{array}
\right)
\end{math}
and 
\begin{math}
C=
\left(
\begin{array}{clrr}
C_1 \\
C_2 \\
C_3 \\
C_4 \\
C_5 \\
C_6 \\
C_7 \\
C_8
\end{array}
\right)
\end{math}
\\
\\
We provide as Supplementary Material a Matlab code (SGHElastoPlastic.m) which calculates $C_1-C_8$ for any set of parameters ($h$, $E$, $\nu$, $\rho$, $\sigma_Y$, $k$, $P_a$, $n$, $a$, $b$ and $P_0$).

\section{Approximate expression for maximum uplift in the elasto-plastic model}\label{AppD}

For large values of the argument $x$, the asymptotic expressions of $kei_0$ and $kei_0$ are \cite[see][p266, equation j]{Timoshenko1959}:
\begin{eqnarray}
ker_0(x)\sim \sqrt{\frac{\pi}{2x}} e^{-x/\sqrt{2}} cos\left(\frac{x}{\sqrt{2}}+\frac{\pi}{8}\right)\\
kei_0(x)\sim -\sqrt{\frac{\pi}{2x}} e^{-x/\sqrt{2}} sin\left(\frac{x}{\sqrt{2}}+\frac{\pi}{8}\right)\\
\end{eqnarray}

Defining $m=\frac{b}{l_e\sqrt{2}}$, the approximate analytical expression for the maximum uplift in the elasto-plastic model, $w_{i,max}$ is given by:

\begin{eqnarray}
& &-\frac{64Dw_{i,max=}}{a^4(P_0-q_0)}=\\ \nonumber
& & \frac{4(P_0+\sigma_Y)}{(q_0-P_0)} \log \left(\frac{b}{a}\right) 
+\frac{q_0+\sigma_Y}{q_0-P_0}
\frac{b^4 \left(64 m^4+384 m^3+960 m^2+1440 m+945\right)}
{a^4 \left(64 m^4+128 m^3+64 m^2-15\right)}
+\\ \nonumber
& & \frac{P_0+\sigma_Y}{q_0-P_0}
\frac{\left(a^2 \left(192 m^4+640 m^3+576 m^2+160 m-45\right)
-4 b^2 \left(64 m^4+256 m^3+384 m^2+400 m+225\right)\right)}
{a^2 \left(64 m^4+128 m^3+64 m^2-15\right)}
\label{ApproxMaxUplift}
\end{eqnarray}

%
%
%
%


%

\bibliographystyle{agufull08}

\end{document}